\documentclass[letterpaper,english,reprint, preprintnumbers, aps, superscriptaddress]{revtex4-1}
\usepackage[T1]{fontenc}
\usepackage[latin9]{inputenc}
\usepackage{color}
\usepackage{amsmath}
\usepackage{amssymb}
\usepackage{graphicx}
\usepackage{esint}

\makeatletter

\providecommand{\tabularnewline}{\\}

\@ifundefined{showcaptionsetup}{}{%
 \PassOptionsToPackage{caption=false}{subfig}}
\usepackage{subfig}
\makeatother

\usepackage{babel}
\begin{document}
\preprint{UC Irvine, October 2024}
\title{Small-Signal Model for Inhomogeneous Helix Traveling-Wave Tubes using
Transfer Matrices}
\author{Robert Marosi}
\author{Kasra Rouhi}
\author{Tarek Mealy}
\affiliation{Department of Electrical Engineering and Computer Science, University
of California, Irvine, Irvine, California, 92697}
\author{Alexander Figotin}
\affiliation{Department of Mathematics, University of California, Irvine, Irvine,
California, 92697}
\author{Filippo Capolino}
\email{f.capolino@uci.edu}

\affiliation{Department of Electrical Engineering and Computer Science, University
of California, Irvine, Irvine, California, 92697}
\date{\today}
\begin{abstract}
We introduce a practical method for modeling the small-signal behavior
of frequency-dispersive and inhomogeneous helix-type traveling-wave
tube (TWT) amplifiers based on a generalization of the one-dimensional
Pierce model. Our model is applicable to both single-stage and multi-stage
TWTs. Like the Pierce model, we assume that electrons flow linearly
in one direction, parallel and in proximity to a slow-wave structure
(SWS) which guides a single dominant electromagnetic mode. Realistic
helix TWTs are modeled with position-dependent and frequency-dependent
SWS characteristics, such as loss, phase velocity, plasma frequency
reduction factor, interaction impedance, and the coupling factor that
relates the SWS modal characteristic impedance to the interaction
impedance. For the multi-stage helix TWT, we provide a simple lumped
element circuit model for combining the stages separated by a sever,
or gap, which attenuates the guided circuit mode while allowing the
space-charge wave on the beam to pass freely to the next stage. The
dispersive SWS characteristics are accounted for using full-wave eigenmode
simulations for a realistic helix SWS supported by dielectric rods
in a metal barrel, all of which contribute to the distributed circuit
loss. We compare our computed gain vs frequency, computed using transfer
matrices, to results found through particle-in-cell (PIC) simulations
and the 1D TWT code LATTE to demonstrate the accuracy of our model.
Furthermore, we demonstrate the ability of our model to reproduce
gain ripple due to mismatches at the input and output ports of the
TWT.
\end{abstract}

\maketitle

\section{\label{sec:introduction}Introduction}

For decades, traveling-wave tubes (TWTs) have been essential for applications
which require high-power, broadband amplification at microwave and
millimeter-wave frequencies. These applications include, but are not
limited to radar, telecommunication, and electronic warfare. Traveling-wave
tube amplifiers are highly reliable, long lasting, and efficient in
their niche high frequency, high bandwidth, and high power applications,
compared to their solid-state power amplifier counterparts \citep{chong2010latest}.
The basic TWT amplifier is composed of a slow-wave structure (SWS)
and a confined linear electron beam which flows in close proximity
to the SWS. The guided modes that propagate in the SWS in the presence
of the electron beam make the TWT a distributed amplifier by converting
the kinetic energy of the electron beam into amplified waves when
both the beam and the guided waves are velocity synchronized \citep{pierce1947theoryTWT,pierce1951waves}.
There are various types of SWS geometries for TWTs, though the one
we consider here is the most common type: the tape helix. Helix-type
TWTs can efficiently amplify waves over bandwidths of one octave and
higher at moderate power levels, primarily limited by their capability
to dissipate heat \citep{paoloni2021millimeter}. Some notable examples
of helix-type satellite TWTs with high efficiencies, long lifetimes,
and moderate bandwidths can be found in Refs. \citep{bodmer1963satellite,chong2010latest,jacquez198644,sauseng1988high,srivastava1999design,wilson1991high}.
Various wide bandwidth helix-type TWTs for applications such as electronic
warfare have also been showcased in Refs. \citep{datta2009analytical,chong2010latest,frisoni1987theoretical,braatz1987100}.

To amplify waves to power levels on the order of 100 Watts without
needing a high-power input signal, a large power gain is required
for such TWTs. Traveling-wave tubes can be made to exhibit large small-signal
power gain by either increasing the Pierce gain parameter, $C$ (e.g.,
increasing the electron beam current), or by increasing the number
of electronic wavelengths, $N$, in the structure (i.e., increasing
the number of unit cells, $N_{\mathrm{c}}$, in the periodic structure)\citep{pierce1951waves}.
However, to keep the TWT stable against backward wave and regenerative
oscillations at frequencies where the small-signal gain is high, it
is sometimes necessary to separate high-gain TWTs into sections, or
stages, separated by severs. The purpose of the sever is to isolate
TWT stages such that the guided electromagnetic (EM) wave cannot propagate
to the next stage but the space-charge wave supported by the electron
beam is allowed to pass and continue to be modulated in the next stage.
This way, the reflected waves at the output stage of the TWT cannot
return all the way to the input of the structure and be re-amplified.
A typical rule-of-thumb for practical helix TWT designs is to have
no more than 20 dB of gain per stage in TWTs to minimize nonlinearities
and minimize the risk of regenerative oscillations \citep[(Ch. 12)]{gilmour1994principles_ch12}.
Furthermore, practical helix TWTs are often inhomogenous in pitch
along the tube length. For example, in order to improve the matching
of the helix circuit to coaxial input/output terminations on the TWT,
pull-turns (i.e. increasing helix pitch at the input/output ends of
the helix) may be added to the tube, as long as the length of the
pitch transition is long with respect to the guided wavelength \citep[(Ch. 12)]{gilmour1994principles}.
Additionally, to increase the available saturation power and enhance
the beam-wave power conversion efficiency of the TWT, pitch tapers
may be added along the length of the helix. The purpose of the taper
is to maintain velocity synchronization between the beam and guided
waves as the mean velocity of the beam is reduced at the end of the
tube, as explained in \citep[(Ch. 10)]{gilmour1994principles} and
in works such as in Refs. \citep{srivastava2000design,ghosh2008improvements,ghosh2009design,alaria2010design,jung2002positive}.

\section{\label{sec:Problem-Statement}Problem Statement}

\begin{figure}
\begin{centering}
\includegraphics[width=0.9\columnwidth]{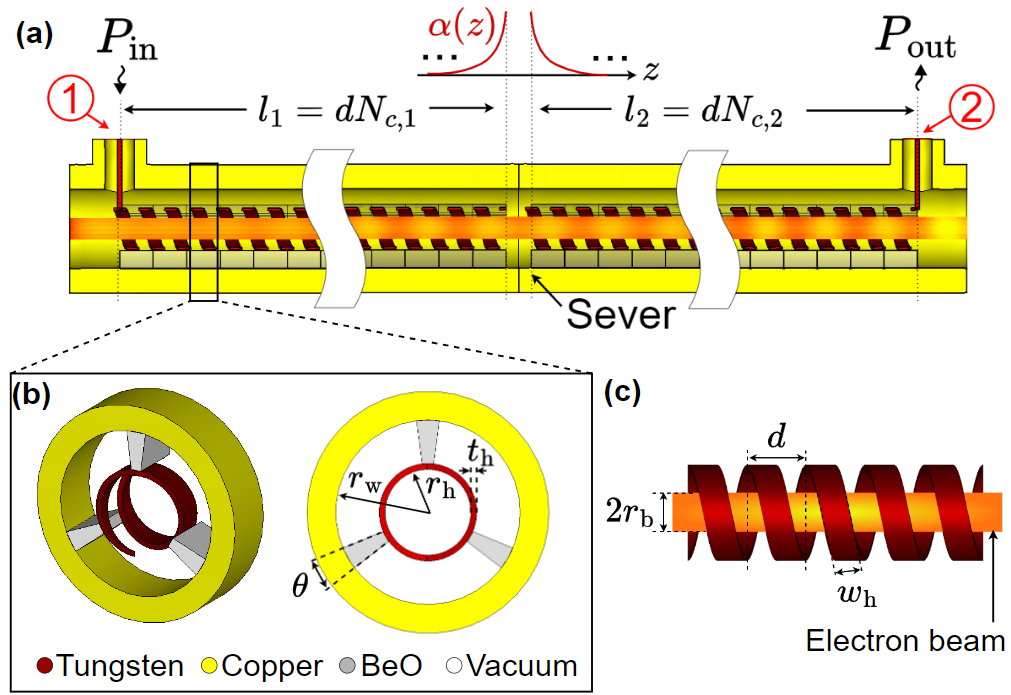}
\par\end{centering}
\caption{(a) TWT interaction region: full-length two-stage SWS with electron
beam, sever and position-dependent attenuation. The electron gun and
collector are not shown here. For simplicity, windows have not been
included at the input or output ports of the structure. Input and
output ports 1 and 2, respectively, are labeled with red circles.
(b) Isometric and front views of a single unit cell of the SWS composed
of a tungsten tape helix supported by three equidistant BeO dielectric
rods within a circular copper waveguide. (c) Horizontal view of the
tape helix SWS with cylindrical electron beam flowing through the
center of the SWS.\label{fig:unit_cell}}
\end{figure}

We develop a simple linear model to predict the small-signal gain
of TWT amplifiers with helix-based SWSs. The linear model works for
interaction regions that consist of either one stage (without a sever)
or two stages seperated by a sever as shown in Fig. \ref{fig:unit_cell}(a).
The sever in a two-stage helix TWT typically consists of a gap between
the helix sections and a position-dependent attenuation on the dielectric
support rods to suppress reflected EM waves. The two-stage helix SWS
geometry has a geometric period $d$ and $N_{\mathrm{\mathrm{c},tot}}=N_{\mathrm{c,1}}+N_{\mathrm{c,2}}$
unit cells, with $N_{\mathrm{c,1}}$and $N_{\mathrm{c,2}}$ cells
on stage 1 and 2, respectively. Furthermore, the helix SWS (either
single-stage or two-stage) can also include position-dependent features
such as phase velocity tapers and/or attenuators. The developed model
accounts for the interaction between the space-charge waves of the
electron beam (also accounting for the space-charge effects) and the
EM wave guided by the SWS. Using this model, it is possible to determine
the RF output power of the EM wave exiting the output port of the
TWT ($P_{\mathrm{out}}$) for a given input RF power ($P_{\mathrm{in}}$)
at the input port of the TWT, assuming that the TWT operates in the
linear (small-signal) regime.

The model described here is based on the generalization of the Pierce
model in \citep{rouhi2021exceptional}. However, unlike the works
of \citep{pierce1947theoryTWT,pierce1951waves,pierce1950traveling1}
and the augmented model in \citep{rouhi2021exceptional}, we make
our model more general in several ways: by considering the dispersive
coupling parameters as discussed in the next section, by modeling
the EM properties the sever gap of the helix SWS using a capacitive
network, and by considering real TWT features that are not necessarily
homogeneous along $z$ (such as pitch tapering and distributed loss
patterning).In Section \ref{sec:Transmission_line_model}, we provide
the analytical formulations for the system and transfer matrices,
based on the generalized Pierce model, that can be used to model discretized
segments of the ``hot'' TWT (i.e., with the electron beam present
inside the SWS) as homogeneous equivalent transmission lines. A more
detailed formulation can be found in \citep{rouhi2021exceptional,rouhi2023parametric}.
Furthermore, we highlight that the model shown in this work is compatible
with TWTs with SWSs other than the helix geometry. For instance, ref.
\citep{rouhi2023parametric} demonstrates this linear model for a
serpentine TWT, albeit without losses or position-dependent features.
However, we restrict our discussion to the single-stage and two-stage
helix-type TWT with arbitrary position-dependent loss distributions.

Using the unit cell dimensions provided in Section \ref{sec:Unit-Cell-Dimensions},
we explain how the interaction impedance and attenuation coefficient
used in our generalized Pierce model are computed using results from
the eigenmode solver of the commercial software CST Studio Suite in
Section \ref{sec:Zp_and_alpha}. In Section \ref{sec:TWT_construction},
we explain how we construct the equivalent transfer matrix for single-stage
and two-stage helix TWTs. Then, in Section \ref{sec:Computation-of-Gain},
we explain how we compute the small-signal gain and evolution of the
state vector for the TWT examples, considering an equivalent lumped
circuit network at the sever and terminations at the input/output
ports of the TWT. Since our model relies on equivalent transmission
lines (TLs) and load impedances, it allows us to observe how mismatches
at the input and output ports of the TWT lead to gain ripple. In Section
\ref{sec:Results}, we show the gain vs frequency and evolution of
the state vector versus position for the TWT examples and compare
the results of our model to those from particle-in-cell (PIC) simulations
and to those from the 1D TWT code LATTE. Finally, in Section \ref{sec:Comparison-to-LATTE},
we explain how our model differs from the TWT code LATTE.

\noindent 

\section{Generalized Pierce Model\label{sec:Transmission_line_model}}
\begin{center}
\begin{figure}
\begin{centering}
\includegraphics[width=0.9\columnwidth]{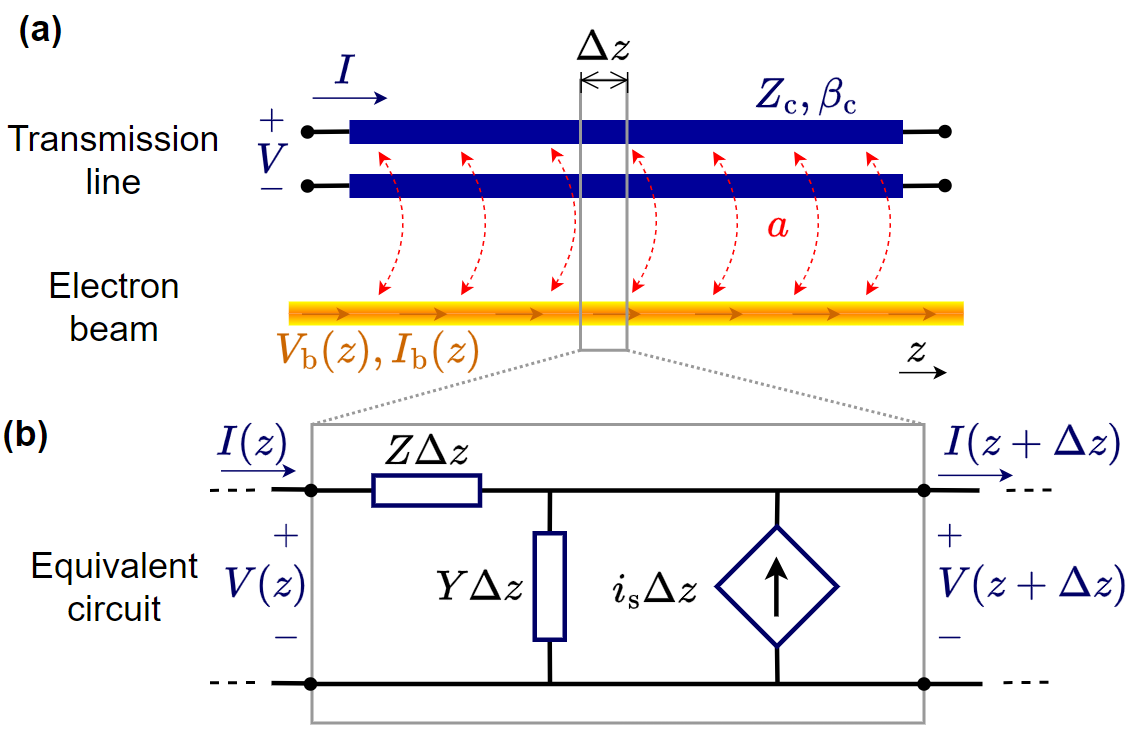}
\par\end{centering}
\caption{(a) Section of an equivalent transmission line of characteristic impedance
$Z_{\mathrm{c}}$ and cold propagation constant $\beta_{\mathrm{c}}$
coupled to an electron beam through the coupling parameter $a$. (b)
Equivalent per-unit-length circuit of transmission line with distributed
impedance $Z$ and admittance $Y$, with current-dependent ``source''
current generator $i_{\mathrm{s}}=-a\left(dI_{\mathrm{b}}/dz\right)$
to account for the effect of the electron beam on the guided EM wave.\label{fig:TL_model}}
\end{figure}
\par\end{center}

To model TWTs that are inhomogeneous along their length, we segment
each TWT stage of length $l$ into $S$ segments of \emph{homogeneous}
equivalent transmission lines (i.e. each segment has a constant attenuation,
phase velocity, and characteristic impedance), similarly to what has
been done in \citep{chernin2012effects}. The homogeneous equivalent
transmission line model for a segment of the TWT stage is illustrated
in Fig. \ref{fig:TL_model}. The homogeneous transmission line segments
for each stage have a fixed length of $\Delta l=l/S$, where $S$
is the number of segments. Note that in this paper, $S$ does not
need to be equal to the number of unit cells in the SWS, $N_{\mathrm{c}}$,
which allows for more flexibility in the modeling of non homgenous
parameters along the SWS. In the prior work by Rouhi et.al. \citep{rouhi2023parametric},
the TWT was partitioned into segments with length equal to the period
of the serpentine waveguide unit cell (i.e., $\Delta l=d$ and $S=N_{\mathrm{c}}$).
Specifically, subdividing the TWT stages by the geometric period of
the SWS, $d$, works well for TWTs with no position-dependent features
along their length. TWTs with pitch tapering and/or loss patterning
sometimes need to be modeled with more transmission line segments
than the number of unit cells (i.e. $\Delta l<d$) in order to adequately
resolve these features (e.g. TWTs with have loss patterning near the
sever that rapidly varies from one unit cell to the next). Table \ref{tab:Parameters}
summarizes the important parameters used to subdivide the TWT stages
and define their physical length in our model (the subscripts 1 and
2 are later introduced to denote the parameters for the first and
second stages in a two-stage TWT, respectively, in Sec. \ref{sec:TWT_construction}).

\noindent 
\begin{table}
\caption{Parameters used to describe the length of each TWT stage and its segments.\label{tab:Parameters}}

\noindent \centering{}%
\begin{tabular}{|c|c|c|}
\hline 
Parameter & Description & Notes\tabularnewline
\hline 
\hline 
$d$ & Geometric period of the unit cell & \tabularnewline
\hline 
$N_{\mathrm{c}}$ & Number of unit cells in a TWT stage & \tabularnewline
\hline 
$l$ & Length of a TWT stage & $l=N_{\mathrm{c}}d$\tabularnewline
\hline 
$S$ & Number of homogeneous segments & \tabularnewline
\hline 
$s$ & Segment index & $s=1,2,...,S$\tabularnewline
\hline 
$\Delta l$ & Length of a segment in a TWT stage & $\Delta l=l/S$\tabularnewline
\hline 
\end{tabular}
\end{table}

Pierce theory is often used to describe the interaction between the
space-charge waves of an electron beam and the guided EM mode using
equivalent transmission lines in the small-signal regime with linearized
equations. In Pierce theory, the TWT system is considered as a one-dimensional
problem, with a guided EM wave on a SWS interacting with a linear
space- and time-varying electron beam confined by a strong axial magnetic
field, such that only electron movement in the axial ($+z$) direction
is considered. For simplicity, we also assume that the axial electric
field of the modes guided by the SWS and acting on the electron beam
is approximately uniform over the cross-section of the electron beam,
as was done in \citep{pierce1947theoryTWT,pierce1951waves,rouhi2021exceptional,pierce1950traveling1}.
The electron beam has an average equivalent kinetic voltage $V_{0}$
and average beam current $-I_{0}$ in the $+z$ direction (using conventional
flow notation, positive values of $V_{0}$ and $I_{0}$ are used in
the remainder of this paper). The average axial velocity of electrons
in the beam is $u_{0}$ and it is approximately related to the average
kinetic voltage as $V_{0}\approx u_{0}^{2}/\left(2\eta\right)$ for
low-voltage beams (i.e. for $V_{0}$ near 10 kV or below \citep{jassem2020theory}),
where $\eta=e/m_{0}=1.7588\times10^{11}$ C/kg is the charge-to-rest-mass
ratio of an electron. Note that this approximation for the electron
kinetic voltage-velocity relation is needed, since the equations of
the generalized Pierce model used here are derived using that approximation
\citep{rouhi2021exceptional}.\textcolor{red}{{} }For higher beam voltages,
it is necessary to account for changes in the effective electron mass
in the derivation of the electron beam equations of \citep{rouhi2021exceptional}.\textcolor{red}{{}
}Development of a generalized Pierce model with relativistic corrections
is planned for a future paper. Hence, we limit our generalized Pierce
model in this work to the non-relativistic regime, as it builds upon
our previous works (which were also formulated using non-relativistic
equations) \citep{figotin2013multi,tamma2014extension,figotin2021analytic,rouhi2021exceptional,rouhi2023parametric}.
The EM-charge modes that exist when the guided EM wave interacts with
the electron beam are sometimes called ``hot'' modes, whereas passive
EM modes guided by the SWS without the electron beam are called ``cold''
modes. For brevity, only the final equations necessary for the Pierce
model are shown in this section. Detailed derivations of the equations
governing the Pierce model may be found in the original Pierce papers
\citep{pierce1947theoryTWT,pierce1951waves,pierce1950traveling1},
or in \citep{rouhi2021exceptional} and its supplementary material.
Here, we augment the generalized Pierce model of \citep{rouhi2021exceptional,rouhi2023parametric}
to include inhomogeneous features over the length of the TWT.

The equivalent transmission line model for a homogeneous TWT segment
is illustrated in Fig. \ref{fig:TL_model}(b). Under the small-signal,
nonrelativistic approximations of the Pierce model, the four differential
equations governing the dynamics of the electron beam and equivalent
transmission line in the phasor domain can be represented in a compact
matrix form at each frequency $\omega$ and longitudinal position
$z$ along the circuit, with implicit $\exp(j\omega t)$ dependence,
$\frac{d}{dz}\boldsymbol{\Psi}(z)=-j\mathbf{\mathbf{\underline{M}_{\mathit{s}}}}\boldsymbol{\Psi}(z)$,
with $z\in\left[(s-1)\Delta l,s\Delta l\right]$, where the subscript
$s=1,...,S$ is the index corresponding to each segment along each
stage of the TWT, the position-dependent $4\times1$ state vector

\noindent 
\begin{equation}
\boldsymbol{\Psi}(z)=\left[\begin{array}{cccc}
V(z), & I(z), & V_{\mathrm{b}}(z), & I_{\mathrm{b}}(z)\end{array}\right]^{\mathrm{T}},\label{eq:state_vec}
\end{equation}

\noindent represents the modulated voltage and current phasors on
both the beam ($V_{\mathrm{b}}$ and $I_{\mathrm{b}}$) and the SWS
($V$ and $I$ representing the guided electric and magnetic fields
along the SWS as equivalent voltages and currents, respectively \citep{marcuvitz1951representation,felsen1994radiation,rouhi2021exceptional}),
with $\mathrm{T}$ as the transpose operator. In our examples, we
define the voltage, $V$, as the potential difference between the
surface of a conducting sheath helix and surrounding metal walls and
we define the current, $I$, as the longitudinal component of current
along the sheath helix. The sheath model we use is explained in detail
in Refs. \citep{kino1962circuit,paik1969design} and in Appendix \ref{sec:helix_zc}.
By using the sheath helix model, we are able to treat each SWS segment
as a uniform transmission line, even when the segment length $\Delta l$
is smaller than the physical period of the unit cell, $d$. The $4\times4$
system matrix $\mathbf{\underline{M}_{\mathit{s}}}$ is homogeneous
for each segment of the TWT

\begin{equation}
\mathbf{\underline{M}_{\mathit{s}}}=\left[\begin{array}{cccc}
0 & k_{\mathrm{c}}Z_{\mathrm{c}} & 0 & 0\\
k_{\mathrm{c}}/Z_{\mathrm{c}} & 0 & -ag & -a\beta_{0}\\
0 & ak_{\mathrm{c}}Z_{\mathrm{c}} & \beta_{0} & \zeta_{\mathrm{sc}}\\
0 & 0 & g & \beta_{0}
\end{array}\right].\label{eq:system_matrix}
\end{equation}

\noindent In Eqn. (\ref{eq:system_matrix}), $k_{\mathrm{c}}=\beta_{\mathrm{c}}-j\alpha_{\mathrm{c}}$
is the ``cold'' complex circuit wavenumber, that also accounts for
losses in the SWS, and $Z_{\mathrm{c}}=V^{+}/I^{+}$ is the characteristic
impedance of a given transmission line segment, $s$, in terms of
the forward voltage and current waves traveling in the positive $z$-direction,
denoted by the + sign (The total voltage and current in the presence
of reflections is the sum of forward-traveling and backward-traveling
waves, i.e. $V(z)=V^{+}(z)+V^{-}(z)$ and $I(z)=I^{+}(z)+I^{-}(z)$,
see Ref. \citep[(Ch. 2)]{pozar2009microwave}). Note that both the
characteristic impedance and propagation constant in the matrix above
can vary from segment to segment due to effects such as pitch tapering
and position-dependent atteunation. For brevity, segment indices are
not included in the subscripts for $k_{\mathrm{c}}$ and $Z_{\mathrm{c}}$.
The definition of voltage and current on the helix SWS is chosen to
yield the net EM power as $P=\frac{1}{2}\mathrm{Re}(VI^{*})$, where
the asterisk ($*$) denotes complex conjugation. Here, we use the
sheath helix model from \citep{kino1962circuit,paik1969design} to
define our helix voltage and current, and by neglecting losses in
the sheath helix, we compute the resultant purely-real helix characteristic
impedance $Z_{\mathrm{c}}$ in Appendix \ref{sec:helix_zc} (small
losses are accounted for only by the complex propagation constant
in our generalized Pierce model). Using a purely-real approximation
for characteristic impedance and a complex propagation constant, is
based on the simplifying assumption that the SWS is a low-loss transmission
line \citep[(Ch. 4)]{collin1960field}\citep[(Ch. 2)]{pozar2009microwave}\citep[(Ch. 5)]{ramo1994fields}.
The real part of the cold circuit wavenumber is related to the cold
phase velocity of a segment (of length $\Delta l=l/S$, as explained
above) of the low-loss SWS, $v_{\mathrm{c}}$, as $\beta_{\mathrm{c}}=\omega/v_{\mathrm{c}}$.
Whereas the imaginary part of the propagation constant, $\alpha_{\mathrm{c}}$
is the attenuation per unit length (in Nepers per meter) for guided
waves on the SWS. For each homogeneous segment of the equivalent transmission
line that represents the SWS, the quantities $k_{\mathrm{c}}Z_{\mathrm{c}}$
and $k_{\mathrm{c}}/Z_{\mathrm{c}}$ in Eqn. (\ref{eq:system_matrix})
are related to the equivalent per-unit-length distributed series impedance
and shunt admittance, as $Z=jk_{\mathrm{c}}Z_{\mathrm{c}}$ and $Y=jk_{\mathrm{c}}/Z_{\mathrm{c}}$
respectively, as illustrated in Fig. \ref{fig:TL_model}(b). Furthermore,
the average electronic phase constant is $\beta_{0}=\omega/u_{0}$,
$g=I_{0}\beta_{0}/(2V_{0})$, and $\zeta_{\mathrm{sc}}=2V_{0}\omega_{\mathrm{q}}^{2}/(\omega I_{0}u_{0})$,
as was also considered in \citep{mealy2019exceptional,rouhi2021exceptional,rouhi2023parametric}.
Note that $i_{\mathrm{s}}$ from Fig. \ref{fig:TL_model} is represented
in Eqn. (\ref{eq:system_matrix}) as $i_{\mathrm{s}}=-a(dI_{\mathrm{b}}/dz)=jagV_{\mathrm{b}}+ja\beta_{0}I_{\mathrm{b}}$.
The frequency-dependent quantity $\omega_{\mathrm{q}}=R_{\mathrm{sc}}\omega_{\mathrm{p}}$
is the reduced plasma angular frequency of the electron beam, which
is proportional to the plasma frequency $\omega_{\mathrm{p}}$ by
the plasma frequency reduction factor $R_{\mathrm{sc}}$. The formulas
used for $\omega_{\mathrm{p}}$ and $R_{\mathrm{sc}}$ are provided
in Appendix \ref{sec:Plasma-Frequency-Reduction}. The frequency-dependent
coupling coefficient $a=(Z_{\mathrm{P}}/Z_{\mathrm{c}})^{1/2}$ is
a factor which relates the actual characteristic impedance of the
EM mode guided by the SWS to the beam-EM mode interaction impedance
(sometimes referred to as Pierce impedance), as derived in \citep{rouhi2023parametric}.
We also show, in Appendix \ref{sec:alt_sys_matrix}, that the system
matrix $\mathbf{\underline{M}_{\mathit{s}}}$ may be transformed into
$\mathbf{\underline{M}_{\mathit{s}}^{\prime}}$ to be in terms of
interaction impedance $Z_{\mathrm{P}}$ only (see Appendix \ref{sec:alt_sys_matrix}),
without dependence on the coupling coefficient $a$ or the characteristic
impedance of the SWS $Z_{\mathrm{c}}$. In the original work by Pierce
\citep{pierce1947theoryTWT,pierce1951waves,pierce1950traveling1},
there was no distinction between these two impedances; $Z_{\mathrm{P}}$
was used as both the characteristic impedance and interaction impedance.
However the equivalent voltages and currents on the transmission line
for this transformed system matrix will also be scaled by $a$; one
must consider the power of the guided mode when using the transformed
model, rather than the helix voltage (and characteristic impedance)
that can be defined in multiple ways \citep{henningsen1955coupling}.
We note that, in this paper, the interaction impedance $Z_{\mathrm{P}}$
obtained by full-wave simulations is only used to determine the coupling
coefficient $a$, unless one uses the alternative system matrix formulation
provided in Appendix \ref{sec:alt_sys_matrix}. Generalized Pierce
models using the coupling strength coefficient $a$ have also been
developed in \citep{figotin2013multi,tamma2014extension,figotin2021analytic,rouhi2021exceptional,rouhi2023parametric},
but unlike those previous works, here we consider the fact that the
TWT can be lossy, dispersive, and inhomogeneous along its length.\textcolor{red}{{}
}Furthermore, we demonstrate in Appendix \ref{sec:Dispersion-Relation-from}
that the fourth-order modal dispersion relation from Pierce theory
\citep{pierce1951waves,pierce1947theoryTWT}\citep[Ch. 2]{pierce1950traveling1}\citep[Ch. 8]{tsimring2006electron}
can be obtained using the system matrix of Eqn. (\ref{eq:system_matrix}),
provided that the characteristic impedance $Z_{\mathrm{c}}$is assumed
equal to the interaction impedance $Z_{\mathrm{P}}$(as is done in
conventional Pierce theory) to make the coupling coefficient $a$
equal to unity. We also note that, unlike with our model, the original
works by Pierce did not consider dispersive parameters (e.g. frequency-dependent
interaction impedance and phase velocity for the modes supported by
the cold SWS) in solving for the modal dispersion relation for the
TWT \citep{pierce1951waves,pierce1950traveling1,pierce1947theoryTWT}.

Realistic SWSs for TWTs are both dispersive and lossy, with position-dependent
features such as pitch tapering and loss patterning. Because of this
fact, the characteristic impedance $Z_{\mathrm{c}}$, coupling factor
$a$, and complex cold propagation constant $k_{\mathrm{c}}$ of the
equivalent uniform transmission line segments can be position dependent
and frequency dependent. The real and imaginary parts of the cold
complex propagation constant $k_{\mathrm{c}}$ and the purely-real
interaction impedance $Z_{\mathrm{P}}$ of the SWS are computed through
full-wave eigenmode simulations with CST Studio Suite (or equivalent
EM simulation software) for lossy and dispersive SWSs, as explained
in Section \ref{sec:Zp_and_alpha}.

Transfer matrices are used to relate the state vector $\boldsymbol{\Psi}_{s}=\boldsymbol{\Psi}\left(s\Delta l\right)$
from one homogenous segment to the next and track its evolution along
the length of the TWT, as illustrated in Fig. \ref{fig:cascadedTmatrices},
where the TWT is subdivided into $S$ segments with segment indices
$s=1,...,S$, and

\begin{equation}
\boldsymbol{\Psi}_{s}=\underline{\boldsymbol{\mathrm{T}}}_{s}\boldsymbol{\Psi}_{s-1}.\label{eq:cascade_T_matrix}
\end{equation}

The transfer matrix $\underline{\boldsymbol{\mathrm{T}}}_{s}$ for
each TWT segment in Eqn. (\ref{eq:cascade_T_matrix}) is computed
from the system matrix in Eqn. (\ref{eq:system_matrix}) using matrix
exponentiation

\begin{equation}
\underline{\boldsymbol{\mathrm{T}}}_{s}=\exp\left(-j\mathbf{\mathbf{\underline{M}_{\mathit{s}}}}\Delta l\right).\label{eq:T_matrix}
\end{equation}

Note that $\boldsymbol{\Psi^{\mathrm{i}}}=\boldsymbol{\Psi}(z=0)$
in Fig. \ref{fig:cascadedTmatrices} corresponds to the excitation
state vector applied to the input-end of the TWT, whereas the state
vector at the output end of the TWT is $\boldsymbol{\Psi^{\mathrm{o}}}=\boldsymbol{\Psi}_{S}$.
Using such transfer matrices, along with appropriate boundary conditions,
we can compute the small-signal gain of non-homogeneous, single-stage
or multi-stage TWTs at each frequency, as we will show later in Section
\ref{sec:Computation-of-Gain}.
\begin{center}
\begin{figure}
\begin{centering}
\includegraphics[width=0.9\columnwidth]{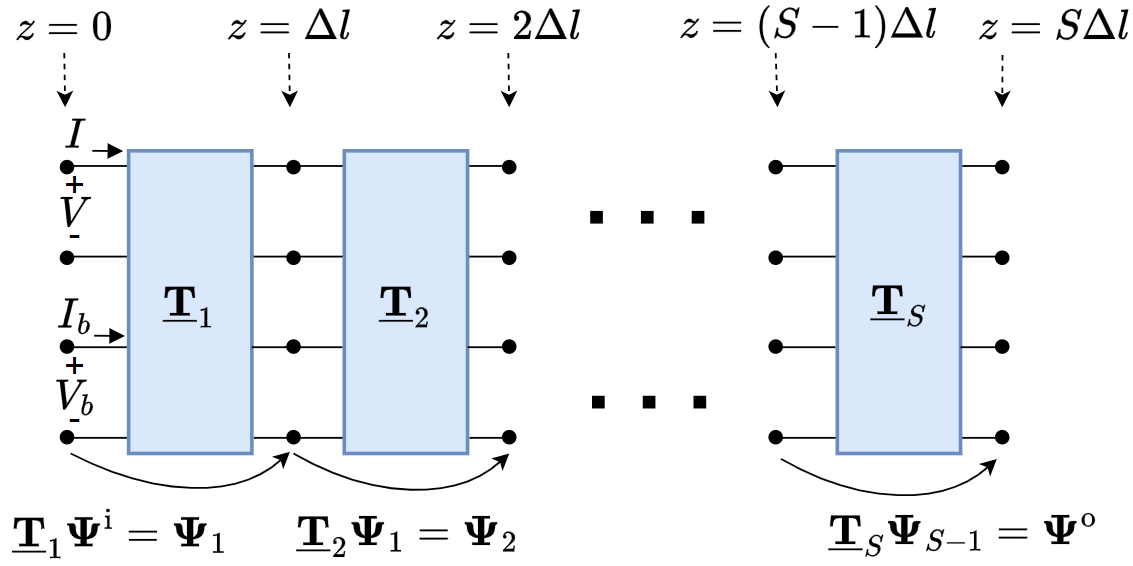}
\par\end{centering}
\caption{Cascading of 4-port transfer matrices to model the spatial evolution
of the state vector with position along the TWT. Each transfer matrix
$\underline{\boldsymbol{\mathrm{T}}}_{s}$ represents a segment of
length $\Delta l$ in the TWT. \label{fig:cascadedTmatrices}}
\end{figure}
\par\end{center}

\section{SWS Unit Cell Dimensions and Parameters \label{sec:Unit-Cell-Dimensions}}

The unit cell used in our single-stage TWT and two-stage TWT examples
is shown in Figs. \ref{fig:unit_cell}(b) and \ref{fig:unit_cell}(c).
The tungsten tape helix (with conductivity $\sigma=1.89\times10^{7}$~S/m)
has a tape thickness $t_{\mathrm{h}}=0.102$~mm, a helix inner radius
$r_{\mathrm{h}}=0.744$~mm, and it is supported by three equidistant
beryllium oxide (BeO) dielectric rods (with $\varepsilon_{\mathrm{r}}=6.53$
and intrinsic $\mathrm{tan}(\delta)=3.58\times10^{-4}$ at 9.3 GHz
and 300 K \citep{daywitt1985complex}, in the ``clean'' regions
of the SWS that have no loss coating). Note however, that the value
of the effective loss tangent in the central region of the finite-length
SWS is position-dependent due to the attenuation patterning. This
attenuation patterning is required to mitigate regenerative oscillations,
as explained in Sec. \ref{sec:TWT_construction}. Furthermore, the
loss tangent is assumed to be constant with respect to frequency in
our full-wave time-domain simulations (for finding the scattering
parameters of the finite-length SWS and for finding the gain of the
TWT in PIC simulations). The dielectric rods subtend angles of $\theta=14.2^{\circ}$
within a circular copper waveguide of inner radius $r_{\mathrm{w}}=1.60$~mm
(with conductivity $\sigma=5.96\times10^{7}$~S/m). Furthermore,
the tape helix has a tape width $w_{\mathrm{h}}=0.51$~mm and helix
pitch (unit cell period) $d=1.04$~mm. The radial thickness of the
copper waveguide is not a critical parameter in full-wave eigenmode
simulations, provided that it is much greater than the skin depth.

\section{Interaction Impedance and Attenuation Coefficient of the SWS\label{sec:Zp_and_alpha}}

Like in the work of \citep{rouhi2021exceptional,rouhi2023parametric},
the per-unit-length transmission line parameters for a homogeneous
``cold'' SWS (i.e. without an electron beam present) can be computed
using frequency-dependent results from full-wave eigenmode simulations.
With the dimensions and materials of the unit cell for the helix SWS
provided in Sec. \ref{sec:Unit-Cell-Dimensions}, we use the finite-element
eigenmode solver of CST Studio Suite to evaluate the cold phase velocity,
$v_{\mathrm{c}}(\omega)$, and interaction impedance, $Z_{\mathrm{P}}(\omega)$,
of the guided mode of the fundamental space harmonic in the cold structure,
like in \citep{kory1998accurate,kory1998effect,aloisio2005analysis}.
However, we also consider the effect of per-unit-length loss, $\alpha$,
in the SWS, which can also be determined from full-wave eigenmode
simulations as described below. In Ref. \citep{rouhi2023parametric},
it was shown that the characteristic impedance of the SWS, regardless
of what voltage and current definitions are used to define it, is
related to the interaction impedance, $Z_{\mathrm{P}}$, by the real-valued,
scalar coupling coefficient $a$, as $Z_{\mathrm{c}}=Z_{\mathrm{P}}/a^{2}$.
We show in Appendix \ref{sec:alt_sys_matrix} that the voltage and
current of the state vector $\boldsymbol{\Psi}(z)$ may be transformed
to put the system matrix $\mathbf{\underline{M}}_{s}$ in terms of
the usual interaction impedance $Z_{\mathrm{P}}$ from Pierce theory,
as also explained in \citep{rouhi2023parametric} (accurate voltage
and current definitions are important when calculating the power of
the EM wave along the TWT using our model, but the calculated gain
will be the same regardless of what characteristic impedance is used,
since it is a ratio of powers). In other words, the model can be used
to obtain correct gain-vs-frequency results for any purely-real and
positive characteristic impedance $Z_{\mathrm{c}}$, as long as the
coupling coefficient $a$ is calculated using this chosen characteristic
impedance $Z_{\mathrm{c}}$ and the proper interaction impedance $Z_{\mathrm{P}}$.
As mentioned in Section \ref{sec:Transmission_line_model}, we compute
the characteristic impedance of the SWS segments analytically by considering
the helix as a lossless anisotropically-conducting sheath, with helix
voltage and current definitions provided in Appendix \ref{sec:helix_zc}.
By considering the SWS as a sheath helix, we are able to track how
the state vector evolves continuously along the SWS length, rather
than from cell-to-cell like was done in \citep{rouhi2023parametric},
though one can also obtain meaningful results with other voltage and
current definitions for the helix SWS.

From Pierce theory, the interaction impedance is defined for the fundamental
Floquet-Bloch spatial harmonic as

\begin{equation}
Z_{\mathrm{P}}\left(\beta_{\mathrm{c}}\right)=\frac{\left|E_{\mathit{\mathrm{z}}}\left(\beta_{\mathrm{c}}\right)\right|^{2}}{2\beta_{\mathrm{c}}^{2}P\left(\beta_{\mathrm{c}}\right)},\label{eq:zpierce}
\end{equation}

\noindent where $\beta_{\mathrm{c}}$ is the phase constant within
the fundamental Brillouin zone, defined here as $\beta_{\mathrm{c}}d/\pi\in\left[-1,1\right]$.
For a helix-type TWT, the interaction impedance is typically evaluated
within the fundamental Brillouin zone, where the beam-wave interaction
is desired to occur. However, for space harmonic SWSs, such as the
serpentine waveguide or coupled-cavity structure, one must consider
higher order space harmonics when determining interaction impedance,
like for the serpentine-type TWT studied in \citep{rouhi2023parametric}.
Furthermore, $\left|E_{z}\left(\beta_{\mathrm{c}}\right)\right|$
is the magnitude of the axial electric field phasor (i.e., the $z$-component),
for a given phase constant, along the center of the SWS where the
electron beam will be introduced, and $P\left(\beta_{\mathrm{c}}\right)$
is the time-average total power flux of the mode propagating through
the SWS \citep[(Ch. 10)]{gewartowski1965principles}. The quantity
$\left|E_{\mathrm{z}}\left(\beta_{\mathrm{c}}\right)\right|$ is calculated
from full-wave eigenmode simulations \citep{mm2017interaction,de2020experimental,marosi2022three}
as 
\begin{equation}
E_{\mathrm{z}}\left(\beta_{\mathrm{c}}\right)=\frac{1}{d}\intop_{0}^{d}E_{\mathrm{z}}\left(z,\beta_{\mathrm{c}}\right)e^{j\beta_{\mathrm{c}}z}dz.
\end{equation}
 Conveniently, the EM energy simulated within the unit cell between
periodic boundaries in the eigenmode solver of CST Studio Suite is
always normalized to 1 Joule for each eigenmode solution. Thus, the
time average power flux may be simply calculated as $P=\left(1~\mathrm{Joule}\right)v_{\mathrm{g}}/d$,
where the group velocity $v_{\mathrm{g}}=d\omega/d\beta_{\mathrm{c}}$
can be determined directly from the dispersion diagram using numerical
differentiation. Note that the eigenmode solver of CST Studio Suite
neglects losses when finding the wavenumbers and EM fields of eigenmodes,
so the electric field and wavenumbers used for computing the interaction
impedance also neglect the effect of losses.

An approximation of the per-unit-length frequency-dependent cold attenuation
coefficient (in Nepers/meter) is obtained from full-wave eigenmode
simulations using the method developed in \citep{rao2010estimation}
as

\begin{equation}
\alpha_{\mathrm{c}}\approx\frac{\omega}{v_{\mathrm{g}}}\frac{1}{2Q},\label{eq:alpha}
\end{equation}

\noindent where $Q$ is the frequency-dependent quality factor obtained
from post-processing steps in the full-wave eigenmode solver of CST
Studio Suite. The quality factor calculated from the eigenmode solver
accounts for time-average volume and surfaces losses in the dielectric
and metal materials as $P_{\mathrm{D}}=\pi f\varepsilon\iiint_{V}\tan(\delta)\left|\mathbf{E}\right|^{2}dv$
and $P_{\mathrm{W}}=\frac{1}{2}\sqrt{\frac{\pi\mu f}{\sigma}}\iint_{S}\left|\mathbf{H}_{\mathrm{tan}}\right|^{2}ds$
respectively, where $f$ is the simulation frequency, $\sigma$ is
the conductivity of the metal, $\varepsilon$ is the permittivity
of the dielectric support rods, $\mathrm{tan}(\delta)$ is the position-dependent
loss tangent of the dielectric support rods, $V$ is the volume of
the dielectric support rods, and $S$ is the surface of each conductor
in these two equations only. The time-average power dissipated in
the unit cell is $P_{\mathrm{loss}}=P_{\mathrm{W}}+P_{\mathrm{D}}$,
and the quality factor is $Q=2\pi fW/P_{\mathrm{\mathrm{loss}}}$,
where $W$ is the total EM energy in the unit cell (for the eigenmode
solver of CST, the fields are normalized such that $W=1$ Joule).
Note that in our full-wave simulations, the loss tangent of the dielectric
support rods is assumed to be independent of frequency. However, the
attenuation coefficient $\alpha_{\mathrm{c}}$ is still frequency-dependent
due to the dispersive fields and surface currents in the SWS. The
conductivity, loss tangent, and relative permittivity of each material
used in our examples is provided in Sec. \ref{sec:Unit-Cell-Dimensions}.
Position-dependent losses are implemented by scaling the loss tangent
of the dielectric support rods as we will explain in detail in Section
\ref{sec:TWT_construction}. Dielectric losses scale linearly with
the loss tangent. However, the attenuation coefficient of Eqn. (\ref{eq:alpha})
ultimately depends on both metal losses and dielectric losses. Thus,
the attenuation coefficient cannot be considered directly proportional
to the loss tangent unless the metal losses are negligibly small with
respect to the dielectric losses.

To implement losses in our generalized Pierce model and compute the
equivalent per-unit length transmission line parameters, we assume
that the ``cold'' guided wavenumber, $k_{\mathrm{c}}=\beta_{\mathrm{c}}-j\alpha_{\mathrm{c}}$,
of the SWS is complex, while using a purely-real characteristic impedance
$Z_{\mathrm{c}}$ that is computed for a lossless sheath helix using
the formulas shown in Appendix \ref{sec:helix_zc}. Since the interaction
impedance $Z_{\mathrm{P}}$ is also purely real by definition, using
a purely-real characteristic impedance conveniently results in a real
and positive value for the coupling coefficient $a$.

The phase velocity, characteristic impedance, and interaction impedance
versus frequency for the uniform-pitch structure shown in Fig. \ref{fig:unit_cell},
with dimensions provided in Sec. \ref{sec:Unit-Cell-Dimensions},
are shown in Fig. \ref{fig:vph_zc}. The frequency- and position-dependent
attenuation coefficient is described in detail in Appendix \ref{sec:Attenuation-Pattern}.
Note that scaling the coupling coefficient, $a$, will primarily shift
the amplitude of the peak gain, whereas scaling the plasma frequency
reduction factor, $R_{\mathrm{sc}}$ (by changing the beam radius,
helix inner radius, or electronic phase constant), will shift the
frequency where the gain peak occurs.

\noindent 
\begin{figure}
\noindent \begin{centering}
\includegraphics[width=0.9\columnwidth]{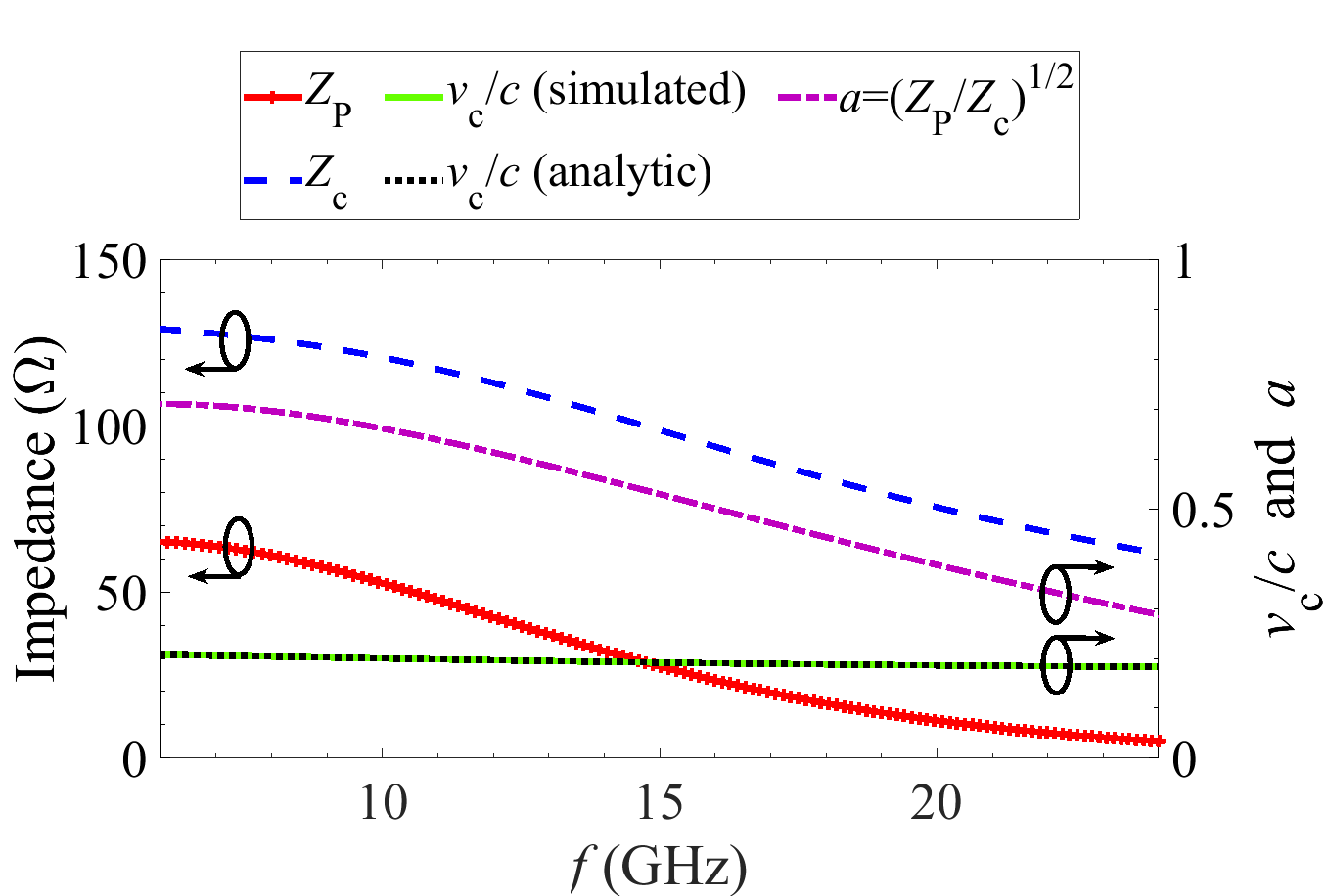}
\par\end{centering}
\caption{Interaction impedance, characteristic impedance, normalized phase
velocity, and coupling coefficient $a$ as a function of frequency
for the helix SWS shown in Fig. \ref{fig:unit_cell} with dimensions
provided in Section \ref{sec:Problem-Statement}. The phase velocity
from the eigenmode solver from CST Studio Suite is compared to the
analytical phase velocity computed using the sheath helix model in
Appendix \ref{sec:helix_zc}, in good agreement.\label{fig:vph_zc}}
\end{figure}

\section{Construction of TWT Transfer Matrices\label{sec:TWT_construction}}

As illustrative examples, we show how we build the transfer matrices
in two simple TWT configurations: (i) a single-stage helix TWT with
uniform pitch (i.e., uniform phase velocity and interaction impedance
with position) and nonuniform loss patterning along its length, and
(ii) a two-stage helix TWT with a sever between stages, uniform helix
pitch on both stages, and nonuniform loss patterning. The commonly-used
Gaussian and exponential loss patterns for each example are defined
analytically in Appendix \ref{sec:Attenuation-Pattern}. The loss
patterns are used to attenuate reflected waves to mitigate the risk
of regenerative oscillations. In practice, position-dependent attenuation
is typically introduced to the helix SWS by coating the dielectric
support rods with carbon via pyrolysis, with a thicker carbon coating
near the center of the sever region \citep{naidu2009three,kumar2008characterization}.
This loss coating can be represented in PIC simulations as an effective
loss tangent in the bulk dielectric material of segmented dielectric
rods. In the examples below, the loss pattern essentially involves
three parameters. The \textit{first}, $\alpha_{\mathrm{min}}$, is
the the minimum attenuation coefficient along the SWS due to intrinsic
dielectric and metal losses. The \textit{second}, $\alpha_{\mathrm{max}}$,
is the maximum attenuation coefficient along the SWS which is obtained
by multiplying the the dielectric loss tangents of the dielectric
support rods by a scaling factor $\tau$ (see Appendix \ref{sec:Attenuation-Pattern}).
Alternatively, one may select a suitable $\alpha_{\mathrm{max}}$
based on our model, and then work backwards to determine the effective
loss tangent corresponding to the desired maximum attenuation coefficient,
provided that the TWT is stable and has an acceptable small-signal
gain in the PIC simulations. This effective loss tangent corresponds
to the lossy regions of the SWS that are heavily coated in carbon.
The \textit{third} parameter is the effective length of the loss pattern
$l_{\alpha}$, which is associated to the full width half maximum
(FWHM) of a Gaussian-shaped loss pattern (for the single-stage TWT)
or the effective length for an exponential-shaped loss pattern (for
the two-stage TWT), corresponding to the distance at which the attenuation
coefficient falls below 1\% of its maximum.

\subsection{Single-Stage TWT Transfer Matrix}

Our first example structure, the single stage TWT, is composed of
a uniform-pitch helix SWS with $N_{\mathrm{c}}=95$ turns (SWS length
$l=N_{\mathrm{c}}d=98.8$~mm), with unit cell dimensions shown in
Fig. \ref{fig:unit_cell} and described in Section \ref{sec:Problem-Statement}.
Note again that, in our examples, the number of unit cells, $N_{\mathrm{c}}$,
is different from the number of segments, $S$, that the TWT stage
is divided into to make the model more flexible. Based on the specific
loss profiles shown in Appendix \ref{sec:Attenuation-Pattern}, we
implement a nonuniform position-dependent Gaussian-shaped loss pattern
in our full-wave PIC simulations by setting a loss tangent for the
dielectric support rods in each individual segment (with segment length
$\Delta l$, as described above) along the length of the SWS, as explained
in Appendix \ref{sec:Attenuation-Pattern}. At the port-ends of the
SWS, the loss pattern for the single-stage SWS has a frequency-dependent
attenuation coefficient of $\alpha_{\mathrm{c}}=\alpha_{\mathrm{min}}$
(calculated using Eqn. (\ref{eq:alpha})) . At the midpoint between
the input and output ports (i.e., at $z=l/2$) the SWS has a frequency-dependent
peak attenuation coefficient $\alpha_{\mathrm{max}}=80\alpha_{\mathrm{min}}$
(determined by scaling the dielectric loss tangent by a dimensionless
factor of $\tau=2000$) , and an effective loss pattern length of
$l_{\alpha}=30$~mm, equal to the FWHM of the Gaussian function,
as is explained in Appendix \ref{sec:Attenuation-Pattern}. The quantities
$\alpha_{\mathrm{min}}$ and $\alpha_{\mathrm{max}}$ correspond the
attenuation coefficients in the ``clean'' and ``deep'' regions
of the loss-coated dielectric support rods, respectively (i.e., the
``clean'' region of the rod has little-to-no carbon coating, whereas
the ``deep'' region has the thickest coating of carbon \citep{goebel1999gain,goebel2000theory}).
In our PIC simulations, using CST Studio Suite, we scale the loss
tangent of the dielectric rods between $\tan\left(\delta\right)$
and $2000\tan\left(\delta\right)$, corresponding to $\alpha_{\mathrm{min}}$
and $\alpha_{\mathrm{max}}$, respectively, using the position-dependent
Gaussian profile mentioned above. The loss tangent values used in
CST Studio Suite are sampled at a single frequency of 9.3 GHz, see
Sec. \ref{sec:Unit-Cell-Dimensions}. In the CST simulations, the
loss tangent at each position is assumed to be constant with respect
to frequency, which is a reasonable approximation over the desired
operating frequency range.

\noindent 
\begin{figure}
\noindent \begin{centering}
\includegraphics[width=0.9\columnwidth]{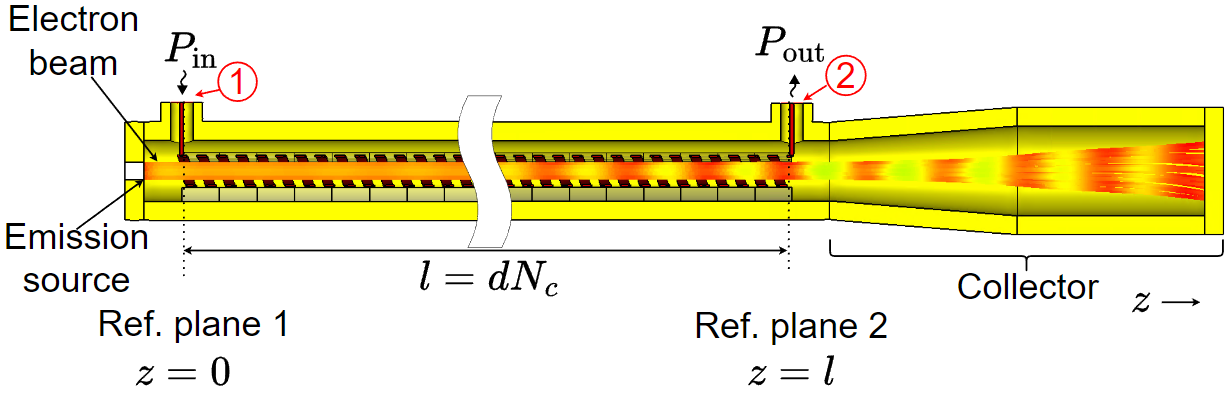}
\par\end{centering}
\caption{Slow-wave structure, emission source, electron beam, collector, and
pin-to-helix transitions for a single-stage helix TWT. Electron bunches
become denser along the SWS length due to beam-wave interaction. Reference
planes for the TL model are indicated at $z=0$ and at $z=l$, between
the two pin-to-helix connection points.\label{fig:singleStageSWS_PIC}}
\end{figure}

The single-stage TWT structure is discretized into $S=200$ homogeneous
segments of length $\Delta l=l/S$. The attenuation coefficient, phase
velocity, and characteristic impedance of the SWS are sampled along
the length of the TWT, where the discrete axial positions along the
TWT are defined using $z=s\Delta l$, where $s=1,...,S$. The location
$z=0$ corresponds to the start of the helical SWS at the input port
of the TWT (port 1), where an input signal is applied, as illustrated
in Fig. \ref{fig:singleStageSWS_PIC}. Using these sampled data, one
can obtain an equivalent system matrix $\mathbf{\underline{M}}_{s}$
defined in Eqn. (\ref{eq:system_matrix}), and an equivalent transfer
matrix $\underline{\boldsymbol{\mathrm{T}}}_{s}$ calculated from
Eqn. (\ref{eq:T_matrix}) for each discrete position along the TWT.
To construct the equivalent transfer matrix of the full SWS coupled
to the electron beam, the transfer matrices of each discrete uniform
segment of the TWT are cascaded using left multiplication as

\begin{equation}
\underline{\boldsymbol{\mathrm{T}}}=\underline{\boldsymbol{\mathrm{T}}}_{S}...\underline{\boldsymbol{\mathrm{T}}}_{2}\underline{\boldsymbol{\mathrm{T}}}_{1}.\label{eq:single-stage-Tmatrix}
\end{equation}

\begin{figure}
\begin{centering}
\includegraphics[width=0.9\columnwidth]{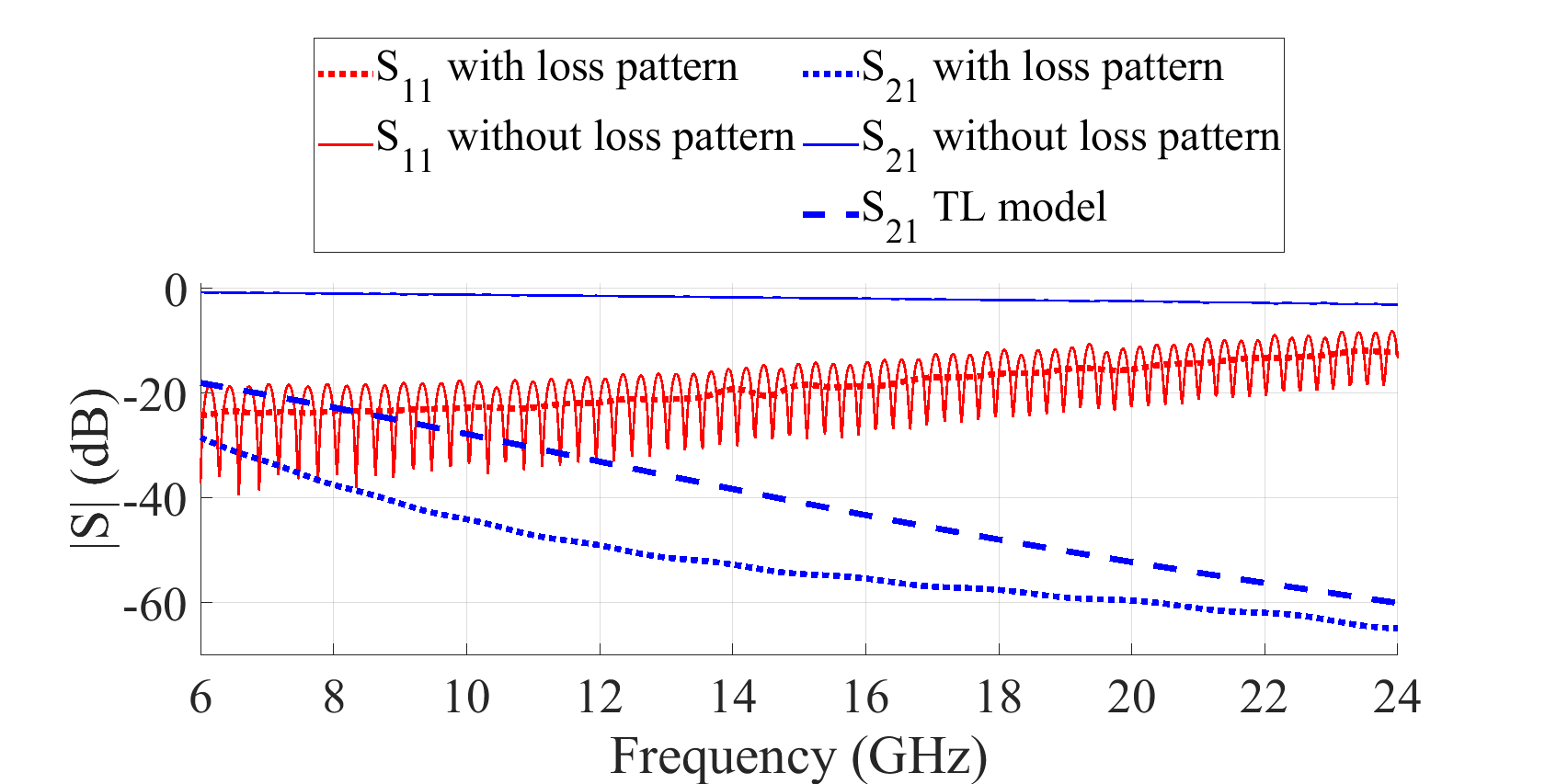}
\par\end{centering}
\caption{Scattering parameters of the single-stage TWT simulated in the time-domain
solver of CST with and without attenuation patterning. The $S_{21}$
of the cold TWT from our TL model (generalized Pierce model with no
beam-wave coupling, i.e., $a=0$, and with a nonzero attenuation coefficient)
is compared to the simulated $S_{21}$ of the SWS with loss patterning.
Note that the $S_{11}$ for the TL model is not shown because it is
on the order of $-300$ dB due to matched terminations used in our
examples (pin-to-helix transitions are not considered). \label{fig:single_stage_cold_sparams}}
\end{figure}

For our example, the scattering parameters of the cold single-stage
TWT, determined from the time-domain solver of CST Studio Suite, indicate
a reflection coefficient below -10 dB and a transmission coefficient
well below -20 dB (due to attenuation) over the chosen operating frequency
range, as shown in Fig. \ref{fig:single_stage_cold_sparams}, allowing
the input and output ports to be well-matched with good isolation
to mitigate backward wave and/or regenerative oscillations. The discrepancies
between the $S_{21}$ of our TL model (with the coupling coefficient
$a=0$ in the system matrix of each segment) and the simulated $S_{21}$(with
loss patterning) in Fig. \ref{fig:single_stage_cold_sparams} are
due to the fact that the TL model does not consider the effect of
pin-to-helix transitions that are present in full-wave simulations.
The TL model describes the interaction region between reference plane
1 that exists just after the pin-to-helix connection at the input
port and reference plane 2 just before the pin-to-helix-connection
at the output port, as illustrated in Fig. \ref{fig:singleStageSWS_PIC}.

\noindent 

\subsection{Two-Stage TWT Transfer Matrix}

\begin{figure}
\noindent \begin{centering}
\includegraphics[width=0.9\columnwidth]{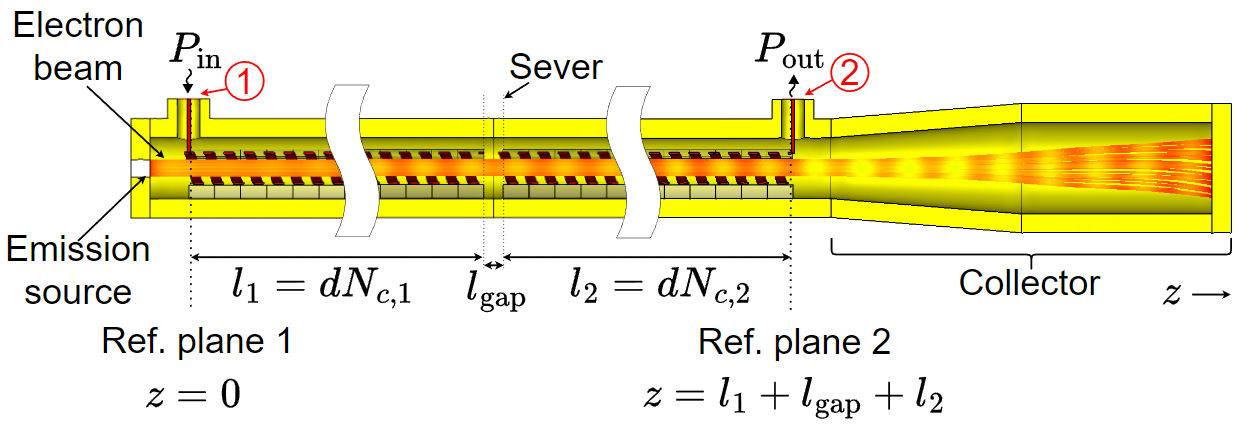}
\par\end{centering}
\caption{Slow-wave structure with sever gap, emission source, electron beam,
collector and pin-to-helix transitions for a two-stage helix TWT.
Electron bunching occurs along the SWS in both stages and bunches
are unaffected by the sever gap. Reference planes for the TL model
are indicated at $z=0$ and at $z=l_{1}+l_{\mathrm{gap}}+l_{2}$,
between the two pin-to-helix connection points.\label{fig:twoStageSWS_PIC}}
\end{figure}

Our second example structure, the two-stage TWT is composed of two
uniform helix stages with $N_{\mathrm{c,1}}=N_{\mathrm{c,2}}=65$
turns ($l_{1}=l_{2}=N_{\mathrm{c,1}}d=N_{\mathrm{c,2}}d=67.6~\mathrm{mm}$)
with the same unit cell dimensions as we used in the previous example,
as illustrated in Fig. \ref{fig:twoStageSWS_PIC}. Note that the subscripts,
1 and 2 are introduced to the length parameters of Table \ref{tab:Parameters}
to indicate the first and second stages, respectively. The two helix
stages are separated by a sever gap length of $l_{\mathrm{gap}}=1~\mathrm{mm}$.
In this drift region (the sever gap illustrated in Fig. \ref{fig:twoStageSWS_PIC}),
there is an outer wall, but no helix or dielectric support rods. The
attenuation pattern is exponential in shape leading into the sever
for each helix stage, as explained in Appendix \ref{sec:Attenuation-Pattern},
where $\alpha_{\mathrm{min}}$ is again calculated using Eqn. (\ref{eq:alpha})
and $\alpha_{\mathrm{max}}=80\alpha_{\mathrm{min}}$ (corresponding
to a scaling of the rod loss tangent by a dimensionless factor of
$\tau=2000$) at the edges of the sever gap (i.e. at positions $z=l_{1}$
and $z=l_{1}+l_{\mathrm{gap}}$). The effective loss pattern length
(the length at which the attenuation decays from its maximum by a
factor of $e^{-5}$, or less than 1\%) on each helix stage is $l_{\alpha}=30$~mm.

\noindent 
\begin{figure}
\noindent \begin{centering}
\includegraphics[width=0.9\columnwidth]{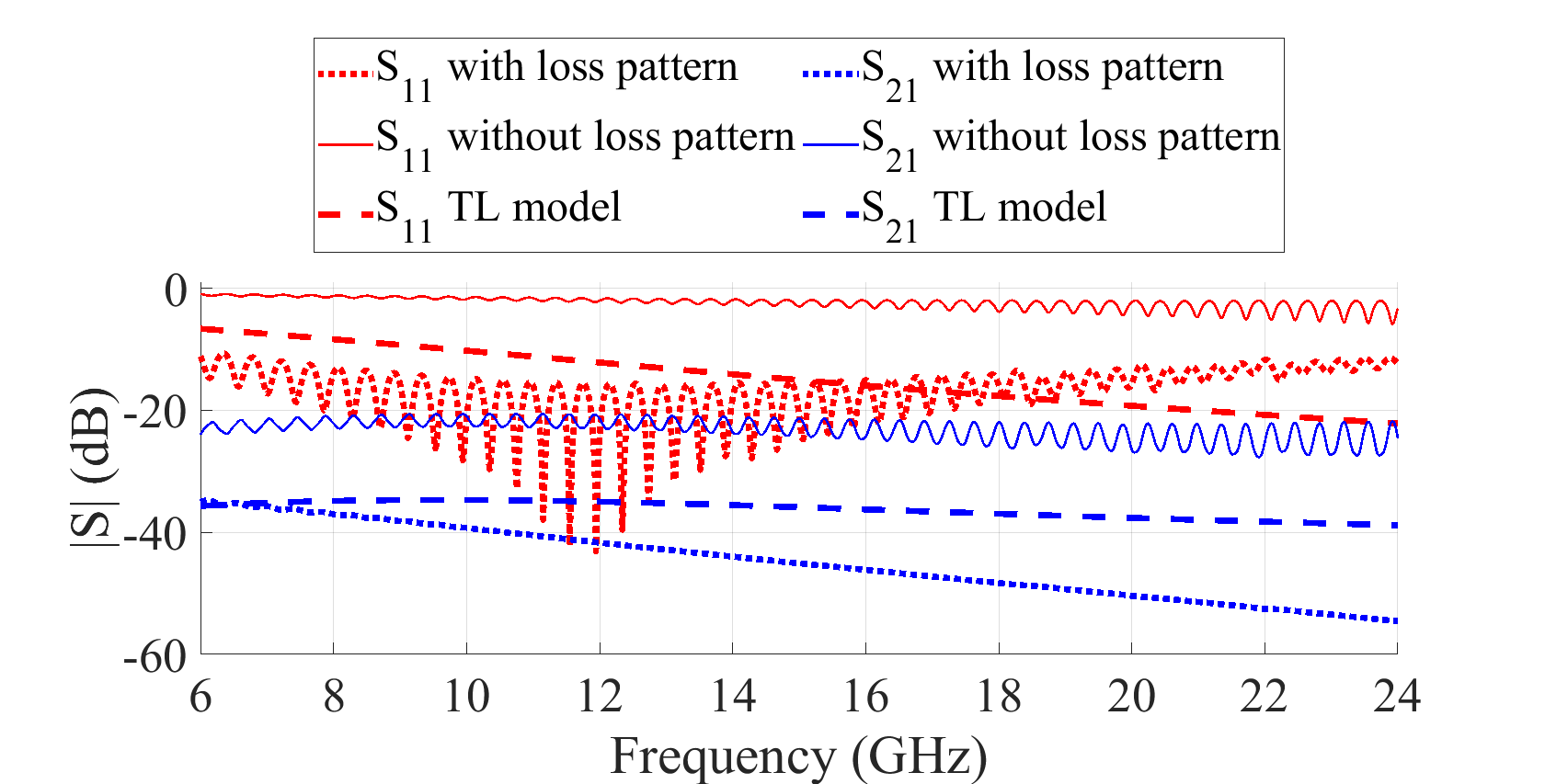}
\par\end{centering}
\caption{Scattering parameters of the two-stage TWT with and without attenuation
patterning. The cold S-parameters of the transmission line model (with
a lumped capacitive network to represent the sever gap, see Appendix
\ref{sec:Sever-Gap-Tmatrix}) are compared to the simulated S-parameters
of the SWS with loss patterning.\label{fig:two_stage_cold_sparams}}
\end{figure}

Each TWT stage is discretized into $S_{1}=S_{2}=200$ homogeneous
segments of lengths $\Delta l_{1}=l_{1}/S_{1}$ and $\Delta l_{2}=l_{2}/S_{2}$,
respectively. The scattering parameters of the two-stage cold TWT
are shown in Fig. \ref{fig:two_stage_cold_sparams}, which indicate
an adequate match at the input and output ports, as well as adequate
isolation between the ports. Again, the discrepancies between the
scattering parameters of our TL model (with the coupling coefficient
$a=0$) and the full-wave simulations of the scattering parameters
(with loss patterning) in Fig. \ref{fig:two_stage_cold_sparams} are
due to the fact that the TL model does not consider the effect of
pin-to-helix transitions that are present in full-wave simulations,
as was also explained in the single-stage example. The input and output
reference planes for the TL model are illustrated in Fig. \ref{fig:twoStageSWS_PIC}
and are adjacent to the pin-to-helix connection points. We obtain
an equivalent system transfer matrix for the input and output stages
of the TWT, denoted by subscripts ST1 and ST2 respectively, by cascading
the homogeneous transfer matrices $\underline{\boldsymbol{\mathrm{T}}}_{1,s}$
and $\underline{\boldsymbol{\mathrm{T}}}_{2,s}$ computed from Eqn.
(\ref{eq:T_matrix}) in the forward direction (left multiplied) for
each discrete position in each TWT stage for the input stage (bunching
stage) as

\begin{equation}
\underline{\boldsymbol{\mathrm{T}}}_{\mathrm{ST1}}=\underline{\boldsymbol{\mathrm{T}}}_{1,S_{1}}...\underline{\boldsymbol{\mathrm{T}}}_{1,2}\underline{\boldsymbol{\mathrm{T}}}_{1,1},
\end{equation}

\noindent and similarly for the output stage (amplification stage)
of the TWT

\begin{equation}
\underline{\boldsymbol{\mathrm{T}}}_{\mathrm{ST2}}=\underline{\boldsymbol{\mathrm{T}}}_{2,S_{2}}...\underline{\boldsymbol{\mathrm{T}}}_{2,2}\underline{\boldsymbol{\mathrm{T}}}_{2,1}.
\end{equation}

However, even though an attenuation pattern is typically used to provide
isolation between TWT stages, helix severs are still non-ideal; they
reflect a large portion of the RF wave (which is typically attenuated
by loss patterning) and allow a small portion of the guided wave to
traverse the sever gap due to capacitive coupling between helices.
Thus, in our model, the RF transmission lines of stage 1 and 2 are
connected using a two-port capacitive pi network with an ABCD matrix
of $\underline{\boldsymbol{\mathrm{T}}}_{\mathrm{gap},\mathrm{c}}$,
which describes the propagation of guided EM waves through the sever
gap. The formulation of $\underline{\boldsymbol{\mathrm{T}}}_{\mathrm{gap},\mathrm{c}}$
is provided in detail in Appendix \ref{sec:Sever-Gap-Tmatrix}. Two
port capacitive networks have also been used to represent center conductor
discontinuities in coaxial transmission lines \citep{whinnery1944coaxial,green1965numerical,dawirs1969equivalent},
which is a similar geometric configuration to the sever problem studied
here. Note that, in the sever gap region, the RF wave is expected
to be quite weak due to heavy distributed attenuation on either side
of the sever gap. The resistive losses that can exist in the sever
gap region are only due to RF wall currents and are neglected here
because they are usualy significantly smaller than the total losses
on either side of the sever gap, and only capacitive coupling betwen
helix stages is considered.

Next, for the beam portion of the sever gap, we use the transfer matrix

\begin{equation}
\underline{\boldsymbol{\mathrm{T}}}_{\mathrm{gap},\mathrm{b}}=\exp\left(-j\mathbf{\underline{M}_{\mathrm{gap,b}}}l_{\mathrm{gap}}\right)
\end{equation}
 that models how the space charge waves along the electron beam behave
in the drift-space of the sever gap, like what was done in \citep{wong2018origin}.
In the sever gap, there should be no coupling between the space charge
wave and the EM guided mode, since there is no helix in the gap. From
the beam equations in \citep{rouhi2021exceptional}, we define the
sever gap system matrix for the beam as

\begin{equation}
\mathbf{\mathbf{\underline{M}_{\mathrm{gap,b}}}}=\left[\begin{array}{cc}
\beta_{0} & \zeta_{\mathrm{sc,gap}}\\
g & \beta_{0}
\end{array}\right],\label{eq:mgap_b}
\end{equation}
where $\zeta_{\mathrm{sc,gap}}=2V_{0}\omega_{\mathrm{q,gap}}^{2}/(\omega I_{0}u_{0})$
uses the reduced plasma angular frequency $\omega_{\mathrm{q,gap}}=R_{\mathrm{sc,gap}}\omega_{\mathrm{p}}$.
This reduced plasma frequency is computed as explained in Appendix
\ref{sec:Plasma-Frequency-Reduction} under the assumption that the
radius of the metal wall in the sever gap is simply $r_{\mathrm{w}}$
instead of $r_{\mathrm{h}}$ . For convenience, the beam and capacitive
gap transfer matrices are combined into one $4\times4$ transfer matrix

\begin{equation}
\underline{\boldsymbol{\mathrm{T}}}_{\mathrm{gap}}=\left[\begin{array}{cc}
\underline{\boldsymbol{\mathrm{T}}}_{\mathrm{gap},\mathrm{c}} & \underline{\mathbf{0}}_{2\times2}\\
\underline{\mathbf{0}}_{2\times2} & \underline{\boldsymbol{\mathrm{T}}}_{\mathrm{gap},\mathrm{b}}
\end{array}\right],
\end{equation}
where $\underline{\mathbf{0}}_{2\times2}$ is a $2\times2$ matrix
containing all zeros. Therefore, the total transfer matrix representing
the two stages of the TWT is computed as

\begin{equation}
\underline{\boldsymbol{\mathrm{T}}}=\underline{\boldsymbol{\mathrm{T}}}_{\mathrm{ST2}}\underline{\boldsymbol{\mathrm{T}}}_{\mathrm{gap}}\underline{\boldsymbol{\mathrm{T}}}_{\mathrm{ST1}}.\label{eq:two-stage-Tmatrix}
\end{equation}

\noindent Next, we calculate the gain of both the single-stage and
two-stage TWT examples by considering the source and load terminations
(i.e., the boundary conditions) at the input and output of the TWT.

\section{Source and Load Terminations\label{sec:Computation-of-Gain}}

The effects of source and load impedances for both the single-stage
and two-stage TWT examples are calculated by writing the equations
for the boundary conditions. The elements of the state vector at the
input and output of each stage are denoted by the superscript $\mathrm{i}$
and $\mathrm{o}$, respectively.

Initially, the single-stage and two-stage TWTs in our TL model are
terminated using matched source and load impedances at each frequency,
i.e., we assume that $Z_{\mathrm{S}}=Z_{\mathrm{L}}=Z_{\mathrm{c}}$.
However, in PIC simulations and in reality, there are small reflections
that occur between the helix SWS and input/output ports (e.g. pin-to-helix
connection, RF windows, and external connections) which can make the
device susceptible to oscillations if the TWT is conditionally stable,
like with microwave transistor circuits \citep[p. 217]{guillermo1984microwave}.
Therefore, we also study the effect of mismatches by considering source
and load terminations that lead to a $-10\,\mathrm{dB}$ reflection
coefficient at both ports of the TWT (i.e., we assume $Z_{\mathrm{S}}=Z_{\mathrm{L}}=Z_{\mathrm{c}}(1+\Gamma)/(1-\Gamma)$
with a $\Gamma=0.1$ that is constant with respect to frequency).
When impedance mismatches are introduced at the TWT ports using our
model, it is possible to observe rippling on the gain versus frequency
and state vector versus position plots, as demonstrated in Sec. \ref{sec:Results}.

\begin{figure}
\noindent \begin{centering}
\includegraphics[width=0.9\columnwidth]{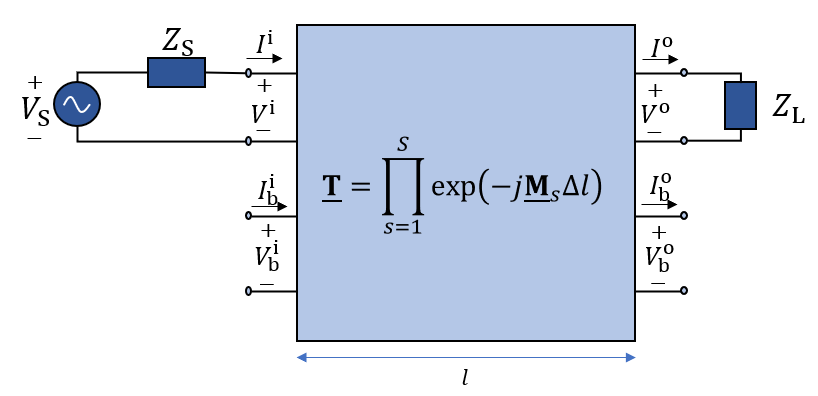}
\par\end{centering}
\caption{Circuit model used to compute the gain of a single-stage TWT of length
$l$, terminated with source and load frequency-dependent impedance
$Z_{\mathrm{S}}$ and $Z_{\mathrm{L}}$, respectively, and driven
with excitation voltage $V_{\mathrm{S}}$. Equivalent transfer matrix
$\underline{\mathbf{T}}$ computed by left-multiplying transfer matrices
for discrete segments of transmission line, of length $\Delta l$.\label{fig:single_stage_gain_model}}
\end{figure}

\begin{figure*}
\noindent \begin{centering}
\includegraphics[width=0.9\textwidth]{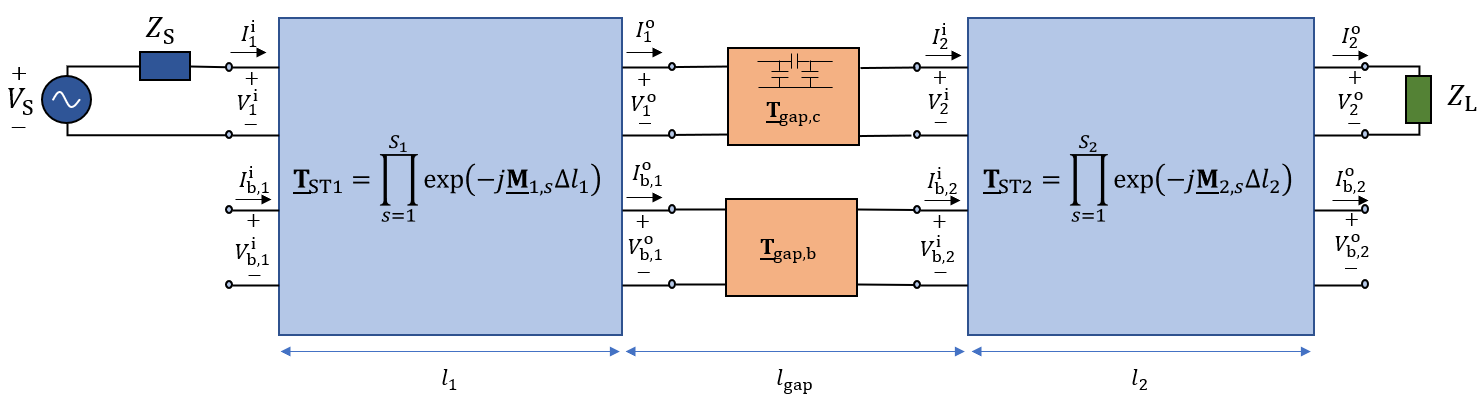}
\par\end{centering}
\caption{Circuit model used to compute the gain of a two-stage TWT. The input
and output stages have lengths $l_{1}$ and $l_{2}$, and the elements
of their state vectors are denoted by the subscript $1$ and 2, respectively.
The sever gap transfer matrices of the TL circuit, $\underline{\boldsymbol{\mathrm{T}}}_{\mathrm{gap},\mathrm{c}}$,
and of the beam, $\underline{\boldsymbol{\mathrm{T}}}_{\mathrm{gap},\mathrm{b}}$,
represent the sever gap of length $l_{\mathrm{gap}}$ that separates
the two helix stages.\label{fig:two_stage_gain_model}}
\end{figure*}

For a single-stage TWT, the initial state of the beam and the input
and output impedances on the ends of the SWS lead to four equations
that must be met at the input and output of the TWT,

\begin{equation}
\left\{ \begin{array}{l}
V_{\mathrm{b}}^{\mathrm{i}}=0\\
I_{\mathrm{b}}^{\mathrm{i}}=0\\
V^{\mathrm{i}}+I^{\mathrm{i}}Z_{\mathrm{S}}=V_{\mathrm{S}}\\
V^{\mathrm{o}}-I^{\mathrm{o}}Z_{\mathrm{L}}=0
\end{array}\right.\label{eq:BC_singleStage}
\end{equation}

\noindent where the ac beam voltage and current are assumed to be
zero at the electron-gun-end of the TWT and an ac voltage, $V_{\mathrm{S}}$
is applied to the input port of the TWT, as illustrated in Fig. \ref{fig:single_stage_gain_model}.
Similarly, for the two-stage TWT, as illustrated in Fig. \ref{fig:two_stage_gain_model},
we have the following equations to represent the boundary conditions,

\begin{equation}
\left\{ \begin{array}{l}
V_{\mathrm{b,1}}^{\mathrm{i}}=0\\
I_{\mathrm{b,1}}^{\mathrm{i}}=0\\
V_{1}^{\mathrm{i}}+I_{1}^{\mathrm{i}}Z_{\mathrm{S}}=V_{\mathrm{S}}\\
V_{2}^{\mathrm{o}}-I_{2}^{\mathrm{o}}Z_{\mathrm{L}}=0
\end{array}\right.\label{eq:BC_twoStage}
\end{equation}

\noindent These two sets of equations, together with $\Psi^{\mathrm{o}}=\underline{\boldsymbol{\mathrm{T}}}\Psi^{\mathrm{i}}$
and $\Psi_{2}^{\mathrm{o}}=\underline{\boldsymbol{\mathrm{T}}}\Psi_{1}^{\mathrm{i}}$
for the single-stage and the two-stage TWTs, respectively, lead to
the $8\times8$ matrix problem $\underline{\boldsymbol{\mathrm{A}}}\mathbf{x}=\mathbf{y}$,
for either TWT under consideration, where 
\begin{equation}
\underline{\boldsymbol{\mathrm{A}}}=\left[\begin{array}{cc}
-\underline{\boldsymbol{\mathrm{T}}} & \underline{\boldsymbol{\mathrm{I}}}_{4\times4}\\
\begin{array}{cccc}
0 & 0 & 1 & 0\\
0 & 0 & 0 & 1\\
1 & Z_{\mathrm{S}} & 0 & 0\\
0 & 0 & 0 & 0
\end{array} & \begin{array}{cccc}
0 & 0 & 0 & 0\\
0 & 0 & 0 & 0\\
0 & 0 & 0 & 0\\
1 & -Z_{\mathrm{L}} & 0 & 0
\end{array}
\end{array}\right],\label{eq:single_stage_A}
\end{equation}

\noindent and $\underline{\boldsymbol{\mathrm{I}}}_{4\times4}$ is
the $4\times4$ identity matrix, and $\underline{\boldsymbol{\mathrm{T}}}$
is the $4\times4$ total transfer matrix of either the single-stage
or the two-stage TWT (given by Eqns. (\ref{eq:single-stage-Tmatrix})
and (\ref{eq:two-stage-Tmatrix}), respectively). Furthermore, the
known vector of this system is

\begin{equation}
\mathbf{y}=\left[\begin{array}{cccccccc}
0, & 0, & 0, & 0, & 0, & 0, & V_{\mathrm{S}}, & 0\end{array}\right]^{\mathrm{T}}.\label{eq:single_stage_b}
\end{equation}

\noindent The unknown $8\times1$ vector $\mathbf{x}$ to be calculated,
that describes the state vectors at both ends of the TWT , is 
\begin{equation}
\mathbf{x}=\left[\begin{array}{cccccccc}
V^{\mathrm{i}}, & I^{\mathrm{i}}, & V_{\mathrm{b}}^{\mathrm{i}}, & I_{\mathrm{b}}^{\mathrm{i}}, & V^{\mathrm{o}}, & I^{\mathrm{o}}, & V_{\mathrm{b}}^{\mathrm{o}}, & I_{\mathrm{b}}^{\mathrm{o}}\end{array}\right]^{\mathrm{T}},
\end{equation}

\noindent for the single-stage TWT, and

\noindent 
\begin{equation}
\mathbf{x}=\left[\begin{array}{cccccccc}
V_{1}^{\mathrm{i}}, & I_{1}^{\mathrm{i}}, & V_{\mathrm{b,1}}^{\mathrm{i}}, & I_{\mathrm{b,1}}^{\mathrm{i}}, & V_{2}^{\mathrm{o}}, & I_{2}^{\mathrm{o}}, & V_{\mathrm{b,2}}^{\mathrm{o}}, & I_{\mathrm{b,2}}^{\mathrm{o}}\end{array}\right]^{\mathrm{T}},
\end{equation}

\noindent for the two-stage TWT.

Solving this $8\times8$ system of equations for the state vector
$\mathbf{x}$ allows us to compute the transducer power gain of the
single-stage TWT as $G=P^{\mathrm{o}}/P^{\mathrm{i}},$where $P^{\mathrm{i}}=V_{\mathrm{S}}^{2}/(8Z_{\mathrm{S}})$
is the \textit{available} RF input power at the input port of the
TWT (or power of the incident wave) and $P^{\mathrm{o}}=\frac{1}{2}\mathrm{Re}\left(V^{\mathrm{o}}I^{\mathrm{o}*}\right)$
is the RF output power delivered to the load at the output port of
the TWT, and $*$ indicates complex conjugation. Similarly, we calculate
the transducer power gain of the two-stage TWT as $G=P_{2}^{\mathrm{o}}/P^{\mathrm{i}},$where
$P_{2}^{\mathrm{o}}=\frac{1}{2}\mathrm{Re}\left(V_{2}^{\mathrm{o}}I_{2}^{\mathrm{o}*}\right)$.

\noindent 

\section{Results: Gain and Evolution of the State Vector\label{sec:Results}}

By applying the boundary conditions described in Sec. \ref{sec:Computation-of-Gain}
(one case with $Z_{\mathrm{S}}=Z_{\mathrm{L}}=Z_{\mathrm{c}}$ for
perfect matching at the input and output port of the TWT at each frequency,
and another case with $Z_{\mathrm{S}}=Z_{\mathrm{L}}=Z_{\mathrm{c}}(1+\Gamma)/(1-\Gamma)$,
corresponding to a constant $-10\,\mathrm{dB}$ mismatch at both ports
at each frequency, where $\Gamma=0.1$), we calculate the gain of
our single-stage and two-stage TWT examples.The gain of the model
described above for the example single-stage and two-stage TWTs is
plotted against frequency in Figs. \ref{fig:single_stage_gain_vs_freq}
and \ref{fig:two_stage_gain_vs_freq}, respectively. The gain of the
model is compared to the gain found from CST PIC simulations and the
open source TWT code LATTE \citep{wohlbier2002multifrequency} using
the same dimensions and parameters as the analytic generalized Pierce
model. The PIC setups for the single-stage and two-stage TWTs are
shown in Figs. \ref{fig:singleStageSWS_PIC} and \ref{fig:twoStageSWS_PIC},
respectively.

The single-stage and two-stage TWTs in our examples are designed to
operate with a dc kinetic equivalent beam voltage of $V_{0}=10.5$~kV
($u_{0}\approx0.2c$) and dc beam current of $I_{0}=50$~mA, with
a beam radius of $r_{\mathrm{b}}=0.46$~mm. For simplicity, we assume
that the SWS in each example is immersed in a strong uniform axial
magnetic field of $B_{\mathrm{z}}=1$~T to confine the electron beam
in our PIC simulations. The electron gun of the TWT in our PIC simulation
is represented as a simple circular area source dc emission. The collector
is a single-stage design, held at the same potential as the walls
of the TWT. A single-tone sinusoidal input signal is applied to port
1 with a constant \emph{accepted} input power of $P_{\mathrm{in}}=-10~\mathrm{dBm}$
at each simulated frequency. If the input termination of the TWT has
an impedance mismatch, the available input power would be slightly
larger than the accepted input power. This is an important detail
to consider when examining the plots of the RF power on the SWS versus
position.

The input power of $-10~\mathrm{dBm}$ was selected to ensure linear
operation for the TWT in PIC simulations, since our model builds on
linearized small-signal equations like in conventional Pierce theory
\citep{pierce1947theoryTWT,pierce1951waves,pierce1950traveling1}.
We have verified in PIC simulations that an input power of $-10~\mathrm{dBm}$
drives the TWT well below the 1 dB gain compression point \citep[Ch. 4]{guillermo1984microwave}
for both TWT examples in this study. Larger input powers on the order
of $30~\mathrm{dBm}$ or higher, for example, may drive the TWT into
the nonlinear saturation regime in reality and in full-wave PIC simulations;
in the saturation regime, our small-signal model cannot accurately
reproduce the results of full-wave PIC simulations and large-signal
TWT codes such as LATTE should be used.

\subsection{Single-Stage TWT}

\begin{figure}
\noindent \begin{centering}
\subfloat[\label{fig:single_stage_gain_vs_freq}]{\noindent \begin{centering}
\includegraphics[width=0.9\columnwidth]{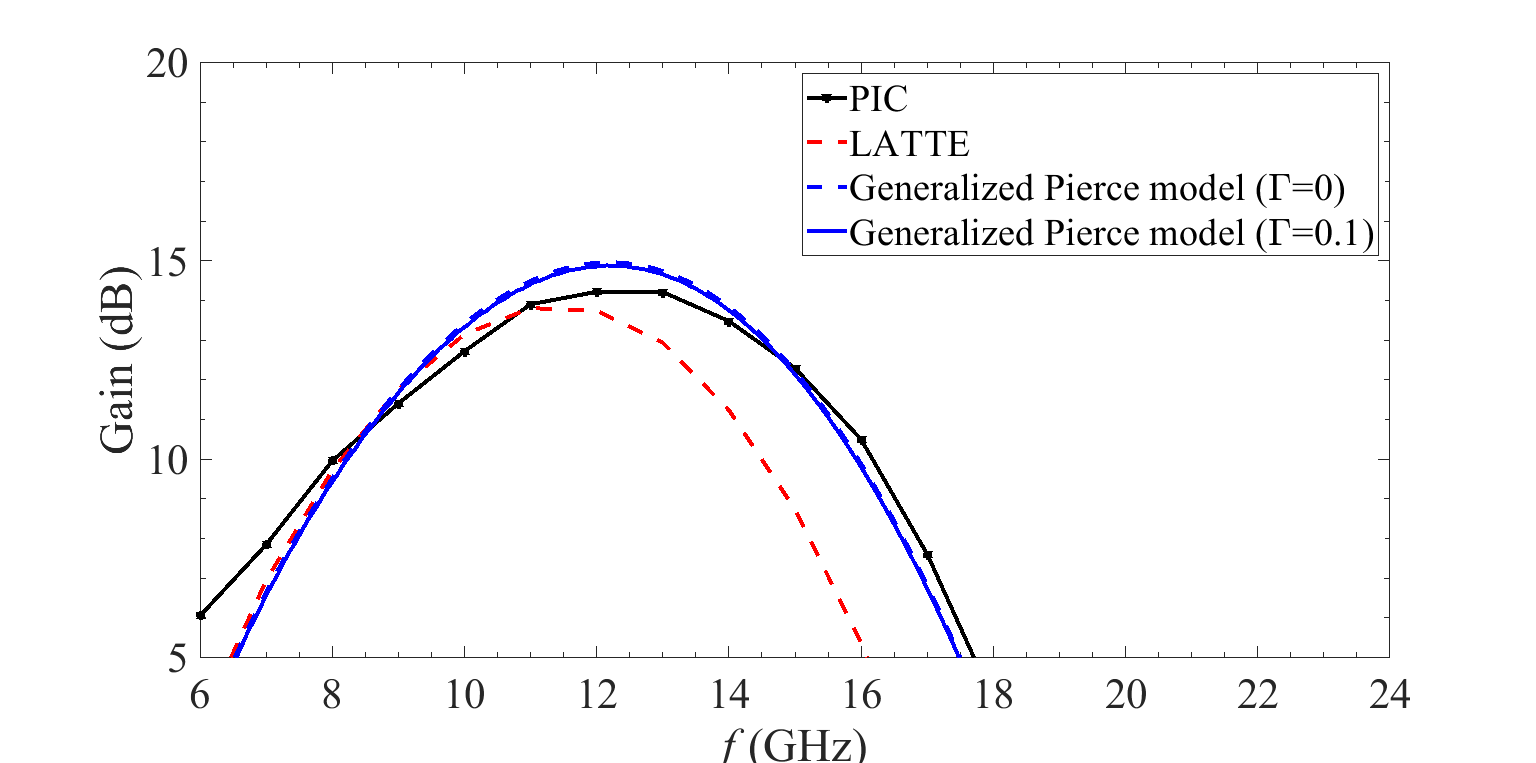}
\par\end{centering}
}\medskip{}
\par\end{centering}
\subfloat[\label{fig:single_stage_state_vec_vs_pos}]{\noindent \begin{centering}
\includegraphics[width=0.9\columnwidth]{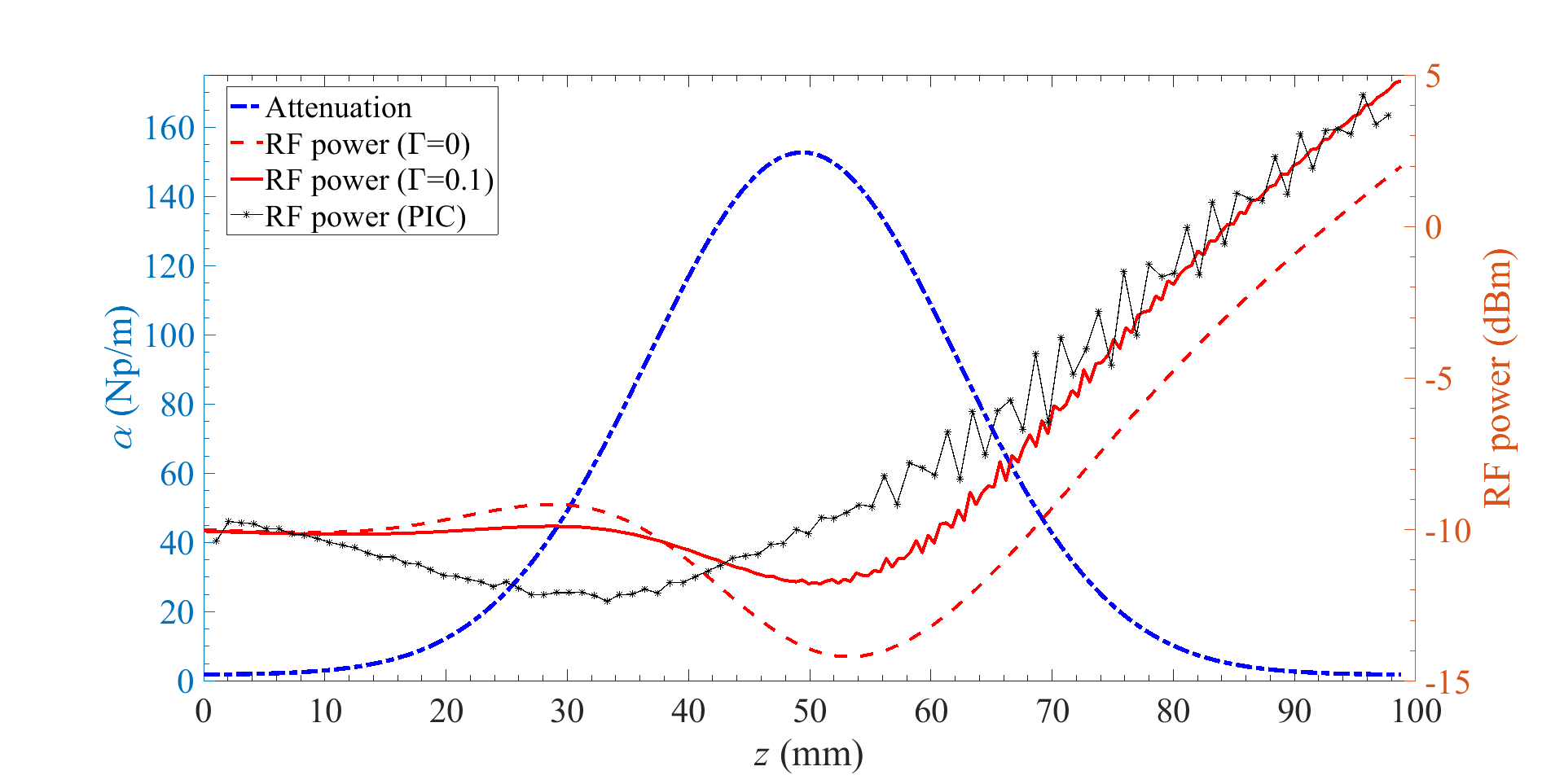}
\par\end{centering}
}

\caption{(a) Gain vs frequency for a single-stage TWT with no mismatches on
both ports ($\Gamma=0$, dashed blue line), and with a $-10\,\mathrm{dB}$
reflection on both ports ($\Gamma=0.1$, solid blue line) calculated
using model described in this paper and compared to PIC and LATTE
simulations. (b) Plot of attenuation coefficient and calculated RF
power along the SWS versus longitudinal position at 12 GHz with and
without a $-10\,\mathrm{dB}$ mismatch (dashed red and solid red lines,
respectively) at the TWT ports using our generalized Pierce model,
in addition to the approximate RF power calculated from field monitors
in PIC simulations (black line with dots).\label{fig:Comparison-of-gain_single_stage}}
\end{figure}

The single-stage TWT example was simulated using the PIC solver of
CST Studio Suite with a simulation duration of 10 ns, approximately
21.3 million mesh cells, and approximately 1.86 million macroparticles.
Also, we use PIC simulations to confirm that the TWT is zero-drive
stable. In this example, the SWS of the TWT consists of $N_{\mathrm{c}}=95$
unit cells. The peak gain of our model (with perfect matching at the
input and output ports) is within approximately 0.9 dB of the peak
gain found using PIC simulations. However, the frequency of the peak
gain is shifted by approximately 0.4 GHz, as shown in Fig. \ref{fig:single_stage_gain_vs_freq}.
Interestingly, the gain from LATTE and our model agree well below
approximately 10 GHz, but the frequency and amplitude of the peak
gain of our model agree better with CST PIC simulations, especially
at frequencies above the frequency of peak gain (above 12 GHz). Using
our model, the state vector was sampled at discrete positions along
the TWT after being cascaded through transfer matrices, as illustrated
in Fig. \ref{fig:cascadedTmatrices}.

The power of the RF wave guided along the circuit, $P(z)=\frac{1}{2}\mathrm{Re}\left(V(z)I^{*}(z)\right)$
, is plotted vs distance at an operating frequency of 12 GHz, with
and without a $-10\,\mathrm{dB}$ mismatch at the TWT ports with our
analytic model, as shown in Fig. \ref{fig:single_stage_state_vec_vs_pos}.
Additionally, we plot the approximate RF power of the guided wave
along the TWT using data from electric field monitors in PIC simulations.
These monitors are placed periodically along the center of the beam
in the TWT (with period $d$), between each unit cell. The approximate
RF power using these field monitors is calculated from Eqn. (\ref{eq:zpierce})
using the steady-state ac magnitude of the longitudinal electric field
(rather than the fundamental spatial harmonic of $E_{z}$) as $P_{\mathrm{PIC}}\approx\left|E_{z}\right|^{2}/\left(2\beta_{c}^{2}Z_{\mathrm{P}}\right)$.
The approximate RF power found from PIC field monitors is reasonably
close to the RF power calculated by our model at the collector-end
of the TWT for two reasons: (i) the fundamental space harmonic of
$E_{z}$ carries more energy than any other space harmonic and is
close to in magnitude to the average $E_{z}$ integrated over the
unit cell period in eigenmode simulations for the helix SWS, (ii)
the growing wave becomes dominant over the other three wave solutions
near the collector-end of the TWT \citep{pierce1951waves,pierce1950traveling1}.
Note that the small dip in RF power on the SWS from 0 mm to 15 mm
can be attributed to the ``launching losses'' of Pierce theory \citep{pierce1951waves,pierce1950traveling1},
where the guided EM wave deposits power to the space charge wave to
modulate it at the start of the tube.

When mismatches are present at the input and output ports of the TWT,
it appears that the power of the RF wave at the output end of the
TWT is larger than the RF output power for the matched TWT in Fig.
\ref{fig:single_stage_state_vec_vs_pos}, but the gain plots with
and without mismatches in Fig. \ref{fig:single_stage_gain_vs_freq}
are quite similar. This apparent discrepancy is due to the fact that
our power gain definition is the transducer power gain, which depends
on the \emph{available} input power for the terminated TWT, rather
than the constant $-10\,\mathrm{dBm}$ accepted (i.e., input) power
used in our examples. When there is a mismatch at the input port of
the TWT, the available RF power $P^{\mathrm{i}}=V_{\mathrm{S}}^{2}/(8Z_{\mathrm{S}})$
will be higher than the accepted RF power. The RF output power delivered
to the load impedance at the end of the TWT, however, is still calculated
as $P^{\mathrm{o}}=\frac{1}{2}\mathrm{Re}\left(V^{\mathrm{o}}I^{\mathrm{o}*}\right)$
and is consistent with the RF power shown in Fig. \ref{fig:single_stage_state_vec_vs_pos}.
Furthermore, in Appendix \ref{sec:Homogeneous-TWT-Example}, we show
a modified example of a spatially homogeneous single-stage TWT with
only intrinsic losses (i.e. losses associated to $\alpha_{\mathrm{min}}$)
to demonstrate that the RF power vs position and gain vs frequency
plots have a similar but not identical profiles to those found through
the conventional three-wave theory used by Pierce for spatially uniform
TWTs (which also neglects reflections and backward wave) \citep[Ch. 9]{pierce1950traveling1}\citep[Ch. 8]{tsimring2006electron}.

\subsection{Two-Stage TWT}

The two-stage TWT example was simulated using the PIC solver of CST
Studio Suite with a simulation duration of 10 ns, approximately 29.9
million mesh cells, and approximately 2.27 million macroparticles.
In this example, the input and output stages of the SWS each consist
of $N_{\mathrm{c,1}}=N_{\mathrm{c,2}}=65$ unit cells. For this example,
the peak gain of our generalized Pierce model (with perfect matching
at the input and output ports) is also within approximately 0.9 dB
of the peak gain found using PIC simulations, with the frequency of
the peak gain for our model approximately 0.7 GHz higher than the
peak gain frequency found from PIC simulations, as shown in Fig. \ref{fig:two_stage_gain_vs_freq}.
The frequency of the peak gain calculated from LATTE was approximately
0.5 GHz below the peak gain obtained through PIC simulations, but
the peak gain agrees within approximately 0.6 dB with our PIC simulations.
Furthermore, the gain-versus-frequency profile of our model consistently
overshoots the gain from PIC simulations at frequencies above 13 GHz,
whereas LATTE consistently undershoots the gain from PIC simulations
at the same frequencies. The state vector was again sampled at discrete
positions along each TWT stage after being cascaded through transfer
matrices and the sever like in the previous example. The power of
the RF wave guided along the circuit is plotted vs distance at an
operating frequency of 12 GHz, as shown in Fig. \ref{fig:two_stage_state_vec_vs_pos}.
In addition, the approximate RF power is plotted using data from electric
field probes in PIC simulations, using the method discussed in the
previous subsection. Plotting the RF power against the position-dependent
attenuation profile of the SWS, we observe that the RF power increases
up to about 70 mm from the input port, becomes extremely small at
the sever gap, and then grows back with distance in the output stage
of the TWT. Unlike the case of the single-stage TWT, the power of
the state vector, $P(z)=\frac{1}{2}\mathrm{Re}\left(V(z)I^{*}(z)\right)$,
for the two-stage TWT exhibits ripples along the length of the first
stage with and without accounting for mismatches at the TWT ports;
this is due to the reflections caused by the capacitive pi network
of the sever shown in Fig. \ref{fig:two_stage_gain_model} that connects
the two stages, as explained in Appendix \ref{sec:Sever-Gap-Tmatrix}.
Gain ripples have also been observed in TWTs with multiple reflections
in \citep{chernin2012effects}. As explained in the above subsection
for the single-stage TWT, the apparent higher RF power at output port
of the TWT when mismatches are present in Fig. \ref{fig:two_stage_gain_model}
is due to our assertion that the accepted (i.e., input) RF power of
the TWT is a constant $-10\,\mathrm{dBm}$, while the available RF
power at the input of the TWT is larger when there is an input impedance
mismatch. Furthermore, for the case of $-10\,\mathrm{dB}$ mismatches
at the TWT ports, a gain ripple of approximately 5 dB near the center
frequency is present. The gain ripple is more significant for the
two-stage TWT example shown due to the higher gain and additional
reflections at the sever that are not completely attenuated. Gain
ripple is generally undesireable for practical TWTs and can be mitigated
in two ways: (a) by improving the match at the helix-to-coax or helix-to-waveguide
transitions at the input and output port of the TWT with various geometries
\citep{henningsen1955coupling} or different dimensions, and (b) significantly
increasing the attenuation leading into the sever gap region to further
mitigate reflected waves.
\noindent \begin{center}
\begin{figure}
\noindent \begin{centering}
\subfloat[\label{fig:two_stage_gain_vs_freq}]{\noindent \begin{centering}
\includegraphics[width=0.9\columnwidth]{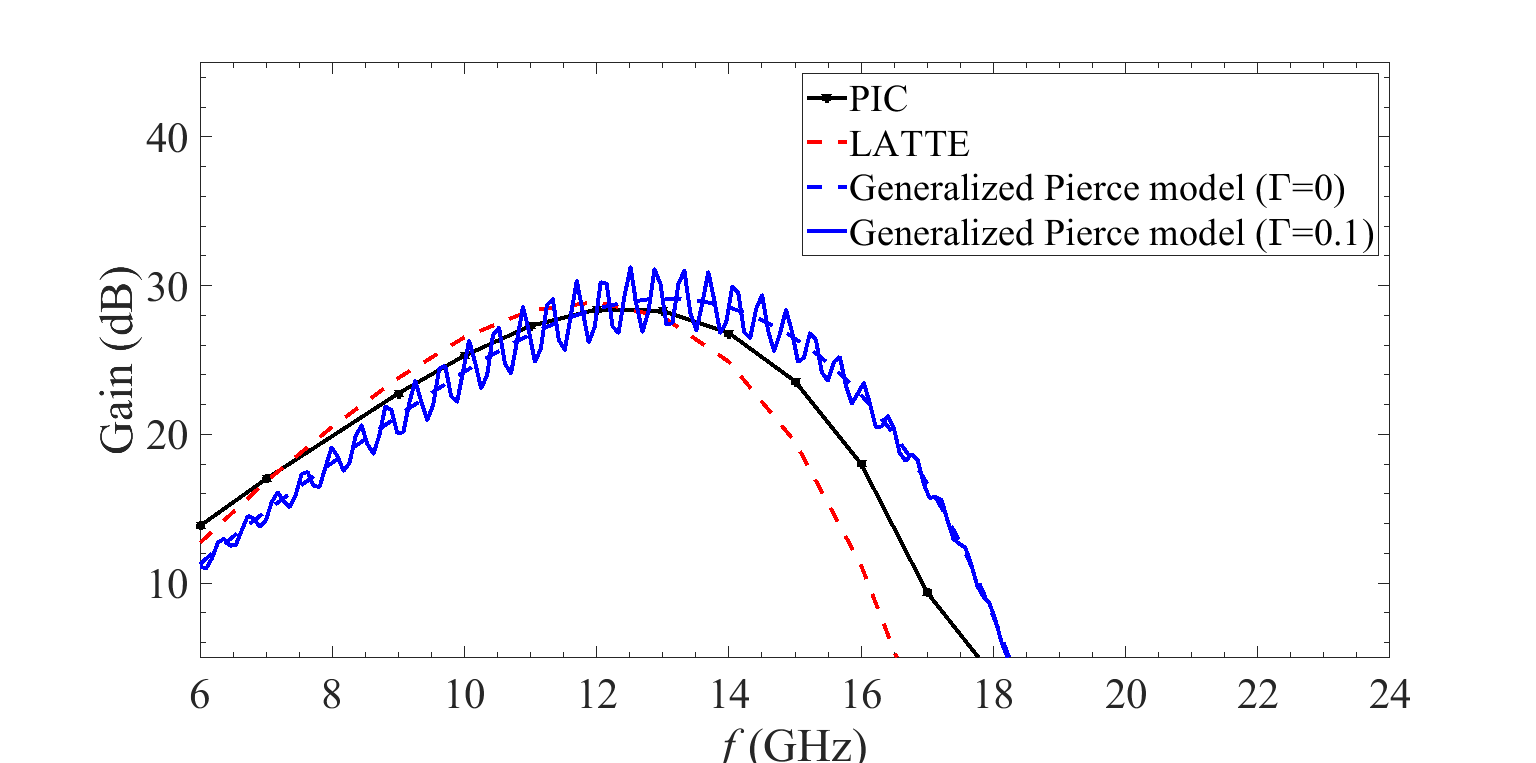}
\par\end{centering}
}\medskip{}
\par\end{centering}
\subfloat[\label{fig:two_stage_state_vec_vs_pos}]{\noindent \begin{centering}
\includegraphics[width=0.9\columnwidth]{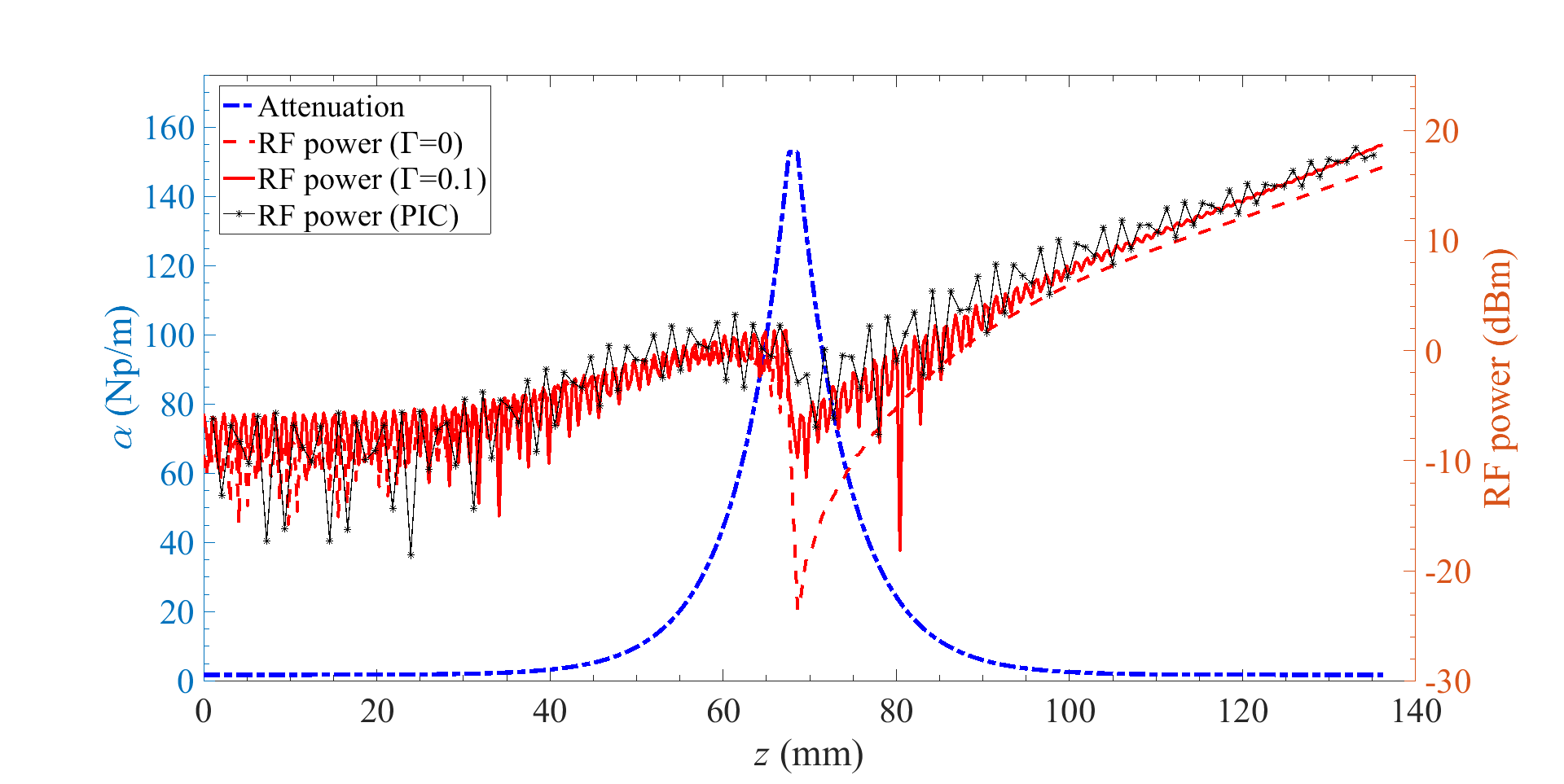}
\par\end{centering}
}

\caption{(a) Gain vs frequency for a two-stage TWT with no mismatches on both
ports ($\Gamma=0$, dashed blue line), and with a $-10\,\mathrm{dB}$
reflection on both ports ($\Gamma=0.1$, solid blue line) calculated
using model described in this paper and compared to PIC and LATTE
simulations. (b) Plot of attenuation coefficient and calculated RF
power along the SWS versus longitudinal position at 12 GHz with and
without a $-10\,\mathrm{dB}$ mismatch (dashed red and solid red lines,
respectively) at the TWT ports using our generalized Pierce model,
in addition to the approximate RF power calculated from field monitors
in PIC simulations (black line with dots).}
\end{figure}
\par\end{center}

\section{Comparison to LATTE\label{sec:Comparison-to-LATTE}}

The TWT code LATTE requires the same frequency- and position-dependent
inputs as our generalized Pierce model, which, aside from the electron
beam parameters, are: interaction impedance, phase velocity, and the
attenuation coefficient. However, LATTE differs from our model in
that it is a Lagrangian-based model; it treats the electron beam as
disks of charge, each interacting individually with the guided EM
mode and other disks of charge \citep{wohlbier2002multifrequency}.
Unlike our model, this feature allows LATTE to also simulate nonlinear
effects such as intermodulation distortion, as well as large-signal
TWT behavior. Many Eulerian- and Lagrangian-based TWT codes also exist
which have been validated and studied in literature \citep{chong2005development,abe2000comparison,srivastava1989one,chernin2001three,kory2001investigation,datta1998nonlinear,dialetis2009accurate,srivastava20032,freund2000three}.
Furthermore, since the average voltage and current of the beam cannot
vary in both the original and our generalized Pierce model, the beam
cannot lose average power to the growing RF wave and net energy conservation
is not possible in such a system. Due to this fact, our model is only
useful in the small-signal regime. However, we also show a simple
demonstration in Appendix \ref{sec:energy-conservation-proof} that,
in the small-signal regime, the time-average power transfer per unit
length between the beam and guided RF wave is nearly conserved under
certain conditions.

However, in the small-signal regime, we believe that the way that
LATTE models the sever gap in multi-stage TWTs is less accurate than
our generalized Pierce model; LATTE ignores impedance mismatches at
the sever gap and instead considers the sever gap as a continuous
length of SWS. This approximation is valid only if the sever gap is
negligibly small with respect to the guided wavelength and if there
is sufficient attenuation to completely suppress reflections which
occur at the sever gap. Furthermore, our transmission line-based model
allows us to consider the effect of unavoidable mismatches at the
input/output terminations of the TWT, in addition to mismatches that
may occur at the sever gap. The code LATTE does not have such capabilities
at this time. In the best case, any mismatches that are present on
the TWT will only lead to gain ripple. In the worse case, the TWT
may become an oscillator. The ability to calculate the effect of such
mismatches in our generalized Pierce model makes it possible to obtain
hot two-port parameters (i.e., the parameters relating equivalent
circuit voltages and currents at the output of the tube to the input
of the tube) for complex TWT designs and, in a future work, predict
the small-signal stability like is done for two-port microwave transistor
amplifiers in Ref. \citep[Ch. 3]{guillermo1984microwave} (in terms
of the parameters $K$ and $\Delta$, there defined, which depend
on active scattering parameters).

\section{Conclusion}

We have showcased a method for determining the small-signal gain of
spatially inhomogeneous, lossy, and dispersive single-stage and multi-stage
TWTs using a generalized Pierce model. The results of our model were
compared against the Lagrangian TWT code LATTE and against full-wave
PIC simulations in CST, with a peak gain that agrees within 1 dB and
within 0.6 GHz with full-wave PIC simulations (when considering matched
terminations at both ends of the TWT). Furthermore, our Pierce-like
model uses the helix dispersive characteristic impedance, $Z_{\mathrm{c}}$,
and a dispersive beam wave coupling factor, $a$, that depends on
the interaction Pierce impedance, $Z_{\mathrm{P}}$. The imperfect
but realistic helix sever in our multi-stage TWT model is represented
using a capacitive pi-network. Furthermore, since our model is based
on equivalent TLs and loads, we are able to consider how impedance
mismatches at the input and output ports of the TWT lead to gain ripple.

Since our model allows for spatially inhomogenous and dispersive TWT
structures to be used, it may serve as a simple and useful tool for
TWT designers to determine suitable and practical TWT lengths, loss
patterns, or pitch profiles for desired small-signal behavior. From
our four-port transfer matrices for both the single-stage and two-stage
TWT models, it may also be possible to determine the ``hot'' two-port
scattering parameters of the TWT and compute the corresponding small-signal
stability factors. Knowledge of these hot scattering parameters may
allow TWT designers to check for zero-drive stability in their TWT
designs before performing time-consuming PIC simulations.
\begin{acknowledgments}
This material is based upon work supported by the Air Force Office
of Scientific Research MURI award number FA9550-20-1-0409 administered
through the University of New Mexico. The authors thank DS SIMULIA
for providing CST Studio Suite, which has been instrumental in this
study. 
\end{acknowledgments}

\section*{}

\section*{Conflict of Interest}

The authors have no conflicts to disclose.

\section*{Data Availability}

The data that support the findings of this study are available from
the corresponding author upon reasonable request.

\appendix

\section{Characteristic Impedance of Guided Modes in a Helix SWS\label{sec:helix_zc}}

Following the work of \citep{kino1962circuit,paik1969design}, we
determine the characteristic impedance of modes guided by a realistic
helix SWS by approximating the tape helix as a lossless, anisotropically
conducting tube, called the sheath helix, which was derived in \citep{osti_4370205}.
Using the sheath model, one may define the transmission line equations
in terms of a voltage, $V$, defined as the potential difference between
the surface of the sheath helix and the conducting wall which encloses
the helix, and a current, $I$, which is defined as the longitudinal
component of the current conducted by the sheath helix. The characteristic
impedance of this sheath helix is also called the longitudinal characteristic
impedance in \citep{paik1969design,henningsen1955coupling}. Paik
also provides a transverse characteristic impedance definition in
\citep{paik1969design} which is proportional to the longitudinal
impedance by the ratio of the group velocity to the phase velocity
$v_{\mathrm{g}}/v_{\mathrm{c}}$, as explained in \citep{paik1969design}.
In this paper, we wish to use the longitudinal characteristic impedance
definition, since the it satisfies the power equation for forward
voltage and current waves $P^{+}(z)=\frac{1}{2}Z_{\mathrm{c}}\left|I^{+}(z)\right|^{2}$,
where the per-unit-length inductance and capacitance are calculated
below as was done in \citep{kino1962circuit,paik1969design}. We will
refer to the longitudinal characteristic impedance as ``characteristic
impedance'' in the remainder of this appendix.

Using these definitions for voltage and current, the transmission
line equations in the phasor domain (with implicit time dependence
as $e^{j\omega t})$ in terms of per-unit-length inductance and capacitance
for a mode guided by a lossless uniform helix SWS are $\frac{dV(z)}{dz}=-j\omega LI\left(z\right)$
and $\frac{dI(z)}{dz}=-j\omega CV\left(z\right),$where the per-unit-length
frequency-dependent inductance $L$ and capacitance $C$ equations
for case 5 in \citep{paik1969design} are shown below. Using the derivation
from \citep{paik1969design}, the characteristic impedance of the
lossless sheath helix may be determined numerically at each frequency
as

\begin{equation}
Z_{\mathrm{c}}=\frac{v_{\mathrm{c}}}{v_{\mathrm{g}}}\sqrt{\frac{L}{C}}.\label{eq:Zc}
\end{equation}
 To compute the characteristic impedance we must compute the quantities
$L$ and $C$ which are given by

\begin{equation}
L=L_{0}\left[1-\kappa_{\mathrm{l}}^{2}\left(\gamma r_{\mathrm{h}},\gamma r_{\mathrm{w}}\right)\right]\label{eq:inductance}
\end{equation}
 and 
\begin{equation}
C=C_{0}\left[1+\left(\theta N_{\mathrm{r}}/(2\pi)\right)\left(\varepsilon_{\mathrm{r}}-1\right)D\left(\gamma r_{\mathrm{h}}\right)\right]\left[1-\kappa_{\mathrm{c}}^{2}\left(\gamma r_{\mathrm{h}},\gamma r_{\mathrm{w}}\right)\right]^{-1},\label{eq:capacitance}
\end{equation}
where $L_{0}=\mu_{0}\beta_{\mathrm{c}}^{2}/\left(2\pi\gamma^{2}\right)\cot^{2}\Psi\left[\mathrm{I_{1}}\left(\gamma r_{\mathrm{h}}\right)\mathrm{K_{1}}\left(\gamma r_{\mathrm{h}}\right)\right]$
and $C_{0}=2\pi\varepsilon_{0}/\left[\mathrm{I_{0}}\left(\gamma r_{\mathrm{h}}\right)\mathrm{K_{0}}\left(\gamma r_{\mathrm{h}}\right)\right]$
are the per-unit-length inductance and capacitance of a sheath helix
of pitch angle $\Psi=\tan^{-1}\left[d/\left(2\pi r_{\mathrm{h}}\right)\right]$
in free space, $\mathrm{I_{1}}$ and $\mathrm{K_{1}}$ are modified
Bessel functions of order 1, $\mathrm{I_{0}}$ and $\mathrm{K_{0}}$
are modified Bessel functions of order 0. Furthermore, the quantities
$\left[1-\kappa_{\mathrm{l}}^{2}\left(\gamma r_{\mathrm{h}},\gamma r_{\mathrm{w}}\right)\right]$
and $\left[1-\kappa_{\mathrm{c}}^{2}\left(\gamma r_{\mathrm{h}},\gamma r_{\mathrm{w}}\right)\right]^{-1}$
in Eqns. (\ref{eq:inductance}) and (\ref{eq:capacitance}) are the
correction factors that account for the metal walls enclosing the
helix, with $\kappa_{\mathrm{l}}^{2}\left(\gamma r_{\mathrm{h}},\gamma r_{\mathrm{w}}\right)=\left[\mathrm{I_{1}}\left(\gamma r_{\mathrm{h}}\right)\mathrm{K_{1}}\left(\gamma r_{\mathrm{w}}\right)\right]/\left[\mathrm{I_{1}}\left(\gamma r_{\mathrm{w}}\right)\mathrm{K_{1}}\left(\gamma r_{\mathrm{h}}\right)\right]$
and $\kappa_{\mathrm{c}}^{2}\left(\gamma r_{\mathrm{h}},\gamma r_{\mathrm{w}}\right)=\left[\mathrm{I_{0}}\left(\gamma r_{\mathrm{h}}\right)\mathrm{K_{0}}\left(\gamma r_{\mathrm{w}}\right)\right]/\left[\mathrm{I_{0}}\left(\gamma r_{\mathrm{w}}\right)\mathrm{K_{0}}\left(\gamma r_{\mathrm{h}}\right)\right]$.
The factor $\left[1+\left(\theta N_{\mathrm{r}}/(2\pi)\right)\left(\varepsilon_{\mathrm{r}}-1\right)D\left(\gamma r_{\mathrm{h}}\right)\right]$
in Eqn. (\ref{eq:capacitance}) accounts for the dielectric loading
of $N_{\mathrm{r}}=3$ rods that have relative permittivity $\varepsilon_{\mathrm{r}}$
that subtend an angle $\theta$, where $D\left(\gamma r_{\mathrm{h}}\right)=\left(\gamma r_{\mathrm{h}}\right)\mathrm{I_{0}}\left(\gamma r_{\mathrm{h}}\right)\mathrm{K_{1}}\left(\gamma r_{\mathrm{h}}\right)$.

The equations (\ref{eq:inductance}) and (\ref{eq:capacitance}) are
substituted into the dispersion relation

\begin{equation}
\beta_{\mathrm{c}}=\omega\sqrt{LC},\label{eq:dispersion_relation_for_zc}
\end{equation}
to solve for the longitudinal propagation constant $\beta_{\mathrm{c}}$
and angular frequency $\omega$ in terms of the radial propagation
constant, $\gamma^{2}=\beta_{\mathrm{c}}^{2}-k_{0}^{2}$, where $k_{0}=\omega\sqrt{\mu_{0}\varepsilon_{0}}$.
Once the relation between the propagation constant and angular frequency
are numerically calculated from Eqn. (\ref{eq:dispersion_relation_for_zc}),
the per-unit-length inductance and capacitance of Eqns. (\ref{eq:inductance})
and (\ref{eq:capacitance}), respectively, can be calculated, along
with the characteristic impedance in Eqn. (\ref{eq:Zc}) that we wish
to find.

Interestingly, we find that the cold phase velocity ($v_{\mathrm{c}}=\omega/\beta_{\mathrm{c}}$)
calculated from the dispersion relation in Eqn. (\ref{eq:dispersion_relation_for_zc})
gives a reasonable approximation of the phase velocities for the real
tape helix structure, as demonstrated in Fig. \ref{fig:vph_zc}, determined
from full-wave eigenmode simulations in CST Studio Suite, even though
it approximates the helix using the sheath model.

\section{Alternative State Vector and System Matrix \label{sec:alt_sys_matrix}}

As explained in the body of this paper and in Sec. \ref{sec:helix_zc},
the choice of voltage, current, and characteristic impedance definitions
for a forward mode guided along the helix must satisfy the power relation
$P=\frac{1}{2}\mathrm{Re}\left[V^{+}I^{+*}\right]$, where in the
cold TL (i.e., without dependent current generator) one has $Z_{\mathrm{c}}=V^{+}/I^{+}$.
Knowledge of this characteristic impedance is useful in the design
of well-matched input or output ports on the TWT. However, in the
original Pierce model, the characteristic impedance of the effective
transmission line is the interaction impedance, $Z_{\mathrm{P}}$
\citep{pierce1947theoryTWT,pierce1951waves,pierce1950traveling1}.
These two impedances are related in this paper by the coupling coefficient
$a$ at each frequency. This coupling coefficient has also seen use
in other works such as \citep{tamma2014extension,rouhi2021exceptional,othman2016theory,othman2016giant,abdelshafy2018electron,figotin2013multi,figotin2021analytic,abdelshafy2021multitransmission,figotin2021exceptional,abdelshafy2022accurate,figotin2022analytic,figotin2023analytic}.

As derived in the appendix of \citep{rouhi2023parametric}, the interaction
impedance and characteristic impedance of the cold SWS are related
through the frequency-dependent coupling coefficient $a$, as

\noindent 
\begin{equation}
a^{2}=\frac{Z_{\mathrm{P}}}{Z_{\mathrm{c}}}.\label{eq:a_squared}
\end{equation}

Using this relation between the helix characteristic impedance and
interaction impedance, one can transform the transmission line voltage
and current of the state vector in Eqn. (\ref{eq:state_vec}) and
system matrix in Eqn. (\ref{eq:system_matrix}) to be in terms of
scaled TL quantities $V^{\prime}(z)=aV(z)$ and $I^{\prime}(z)=I(z)/a$
that maintain the average power definition $P=\frac{1}{2}\mathrm{Re}\left[V(z)I(z)^{*}\right]=\frac{1}{2}\mathrm{Re}\left[V'(z)I'(z)^{*}\right]$,
where $a$ is the same coupling coefficient that relates interaction
impedance to characteristic impedance. Looking at transformed forward
propagating wave, the TL characteristc impedance is the same as the
interaction (or Pierce) impedance, i.e., $V^{\prime+}/I^{\prime+}=Z_{\mathrm{P}}$.
By making this transformation, the state vector and system matrix
are represented (with transmission line segment subscript $s$ suppressed)
as

\begin{equation}
\partial_{z}\boldsymbol{\Psi}^{\prime}(z)=j\mathbf{\mathbf{\underline{M}^{\prime}}}\boldsymbol{\Psi}^{\prime}(z),
\end{equation}

\noindent where the transformed state vector is

\begin{equation}
\boldsymbol{\Psi}^{\prime}(z)=\left[\begin{array}{cccc}
V^{\prime}(z), & I^{\prime}(z), & V_{\mathrm{b}}(z), & I_{\mathrm{b}}(z)\end{array}\right]^{\mathrm{T}},
\end{equation}

\noindent and the transformed system matrix may be expressed in terms
of interaction impedance rather than characteristic impedance as $Z_{\mathrm{P}}=a^{2}Z_{\mathrm{c}}$.

\begin{equation}
\mathbf{\underline{M}^{\prime}}=\left[\begin{array}{cccc}
0 & k_{\mathrm{c}}Z_{\mathrm{P}} & 0 & 0\\
k_{\mathrm{c}}/Z_{\mathrm{P}} & 0 & -g & -\beta_{0}\\
0 & k_{\mathrm{c}}Z_{\mathrm{P}} & \beta_{0} & \zeta_{\mathrm{sc}}\\
0 & 0 & g & \beta_{0}
\end{array}\right].
\end{equation}

We find that, with this transformation, the coupling coefficient is
suppressed from the matrix and one can use $Z_{\mathrm{P}}$ as the
characteristic impedance with the transformed TL parameters $V^{\prime}$
and $I^{\prime}$. This alternate formulation for the system matrix
is useful, since the interaction impedance can be readily found for
a realistic helix SWS using full-wave eigenmode simulations and postprocessing
as in Eqn. (\ref{eq:zpierce}), whereas it may be difficult to define
and determine the equivalent voltage and current and the associated
characteristic impedance in a realistic tape helix SWS. Furthermore,
this alternate system matrix formulation more closely matches the
original works by Pierce \citep{pierce1947theoryTWT,pierce1951waves,pierce1950traveling1}.
The transformed system matrix $\mathbf{\underline{M}^{\prime}}$ may
be used in place of the system matrix $\mathbf{\mathbf{\underline{M}}}$
shown in Eqn. (\ref{eq:system_matrix}) for segments of homogeneous
transmission lines. Furthermore, the power gain and ``hot'' modal
dispersion diagram (for homogeneous structures) of TWTs modeled using
the transformed system matrix $\mathbf{\underline{M}^{\prime}}$ will
be the same as those found using the original system matrix $\mathbf{\underline{M}}$,
based on the fact that $P=\frac{1}{2}\mathrm{Re}\left[V(z)I(z)^{*}\right]=\frac{1}{2}\mathrm{Re}\left[V'(z)I'(z)^{*}\right]$.
The only thing that differs in using $\boldsymbol{\Psi}(z)$ or $\boldsymbol{\Psi}^{\prime}(z)$
is that the \textit{matching terminations} applied to the input and
output of the SWS are either $Z_{\mathrm{c}}$ or $Z_{\mathrm{P}}$,
respectively. This does not mean that the TWT ports should terminated
in reality using one impedance or the other, it means that the definition
of characteristic impedance of the equivalent transmisison line and
source and load impedances must be consistent with the chosen voltage
and current definitions. For helix-based SWSs, the source and load
impedances $Z_{\mathrm{S}}$ and $Z_{\mathrm{L}}$ that match the
SWS' actual characteristic impedance are close to $Z_{\mathrm{c}}$
(using the sheath helix model to analytically approximate the characteristic
impedance and modal dispersion of a tape helix, see Appendix \ref{sec:helix_zc}).
When using the transformed TL parameters $V^{\prime}$ and $I^{\prime}$,
the definition of impedance changes by a factor $a^{2}$ and, in this
new basis, the transformed source and load impedances are $a^{2}Z_{\mathrm{S}}$
and $a^{2}Z_{\mathrm{L}}$, that match with the characteristic impedance
$Z_{\mathrm{P}}$ of the SWS. The method described in this paper can
be used also for unmatched loads, with the mismatch properly taken
into account via the boundary conditions in Eqns. (\ref{eq:BC_singleStage})
or (\ref{eq:BC_twoStage}) for the single-stage or two-stage TWT,
respectively, as shown in Sec. \ref{sec:Results}.

We note that in the body of this paper, we used the notation with
$V$ and $I$ described in Sec. \ref{sec:Transmission_line_model},
and the associated frequency-dependent characteristic impedance $Z_{\mathrm{c}}$
for the helix SWS. We also evaluated the frequency-dependent interaction
impedance $Z_{\mathrm{P}}$ using results from the eigenmode solver
of CST Microwave Studio, as described in Sec. \ref{sec:Zp_and_alpha}
and shown in Fig. \ref{fig:vph_zc}. Using the calculated, frequency-dependent,
characteristic impedance $Z_{\mathrm{c}}$ and the frequency-dependent
interaction impedance $Z_{\mathrm{P}}$ from full-wave eigenmode simulations,
we determined the frequency-dependent coupling coefficient $a$ through
the relation in Eqn. \ref{eq:a_squared}.

\section{Attenuation Pattern\label{sec:Attenuation-Pattern}}

For a single-stage helix TWT, adiabatic position-dependent loss patterning
is added to the dielectric support rods (via carbon coating) to attenuate
any reflected waves that can occur due to imperfect matching at the
input and output ports of the TWT. Thus, we represent this loss patterning
in full-wave PIC simulations by scaling the bulk loss tangent of the
dielectric rods at each position along the structure, with the same
position-dependent profile as our desired attenuation coefficient
used in our theoretical model. In a single-stage TWT, this attenuation
pattern is typically Gaussian in shape, with a peak attenuation halfway
between the input and output ports, as shown in Fig. \ref{fig:single_stage_state_vec_vs_pos}.
This Gaussian profile is defined (in Np/m) as

\begin{equation}
\alpha_{\mathrm{c}}(z)=\Delta_{\alpha}e^{-\left(z-l/2\right)^{2}/\left(2\sigma^{2}\right)}+\alpha_{\mathrm{min}},
\end{equation}
where $\Delta_{\alpha}=\alpha_{\mathrm{max}}-\alpha_{\mathrm{min}}$,
$\alpha_{\mathrm{max}}$ is the peak attenuation coefficient in the
center of the SWS corresponding to the region where the loss tangent
of the dielectric support rods is at its maximum, $\alpha_{\mathrm{min}}$
is the attenuation coefficient at the input or output ends of the
SWS where the loss tangent of the dielectric rods is at its minimum,
$l$ is the total length of the single-stage SWS, $\sigma=l_{\alpha}/\left(2\left[2\ln(2)\right]^{1/2}\right)$,
and $l_{\alpha}$ (effective loss length) is the FWHM of the Gaussian
profile.

Severs are typically implemented in multi-stage helix TWTs by creating
a physical gap between stages, where both helices at the gap are left
unterminated or short-circuited to the barrel of the TWT. Near the
sever, adiabatic, position-dependent loss patterning is again added
to the dielectric support rods to strongly attenuate reflected waves
at the open/short circuited ends of the sever in the same manner described
for single-stage TWTs. In multi-stage TWTs, this loss patterning is
typically exponential in shape leading to either end of the sever,
as shown in Fig. \ref{fig:two_stage_state_vec_vs_pos}. We represent
the position-dependent attenuation coefficient (in Np/m) as an exponential
decaying function from each end of the sever towards the input/output
port with a piecewise function as

\begin{equation}
\alpha_{\mathrm{c}}(z)=\begin{cases}
\Delta_{\alpha}e^{(z-l_{1})\frac{5}{l_{\alpha}}}+\alpha_{\mathrm{min}} & 0\leq z<l_{1}\\
\Delta_{\alpha}e^{-(z-l_{\mathrm{1+g}})\frac{5}{l_{\alpha}}}+\alpha_{\mathrm{min}} & l_{\mathrm{1+g}}\leq z<l_{\mathrm{tot}}\\
0 & l_{1}\leq z<l_{\mathrm{1+g}}
\end{cases}
\end{equation}
where for the two-stage TWT, $l_{\alpha}$ is the effective length
of the attenuation pattern on each stage, the distance at which the
attenuation decays to approximately 99\% from its peak value. Furthermore,
$l_{1}$ and $l_{2}$are the lengths of the first and second stages
of the TWT, respectively, $l_{\mathrm{1+g}}=l_{1}+l_{\mathrm{gap}}$
, $l_{\mathrm{\mathrm{gap}}}$is the length of the sever gap between
TWT stages, and $l_{\mathrm{tot}}=l_{1}+l_{\mathrm{gap}}+l_{2}$ is
the total length of the SWS in a two-stage TWT. Between the two circuit
stages, in the sever gap, there is only vacuum.

\begin{figure}
\noindent \centering{}\subfloat[\label{fig:alpha_min_ratio}]{\noindent \begin{centering}
\includegraphics[width=0.9\columnwidth]{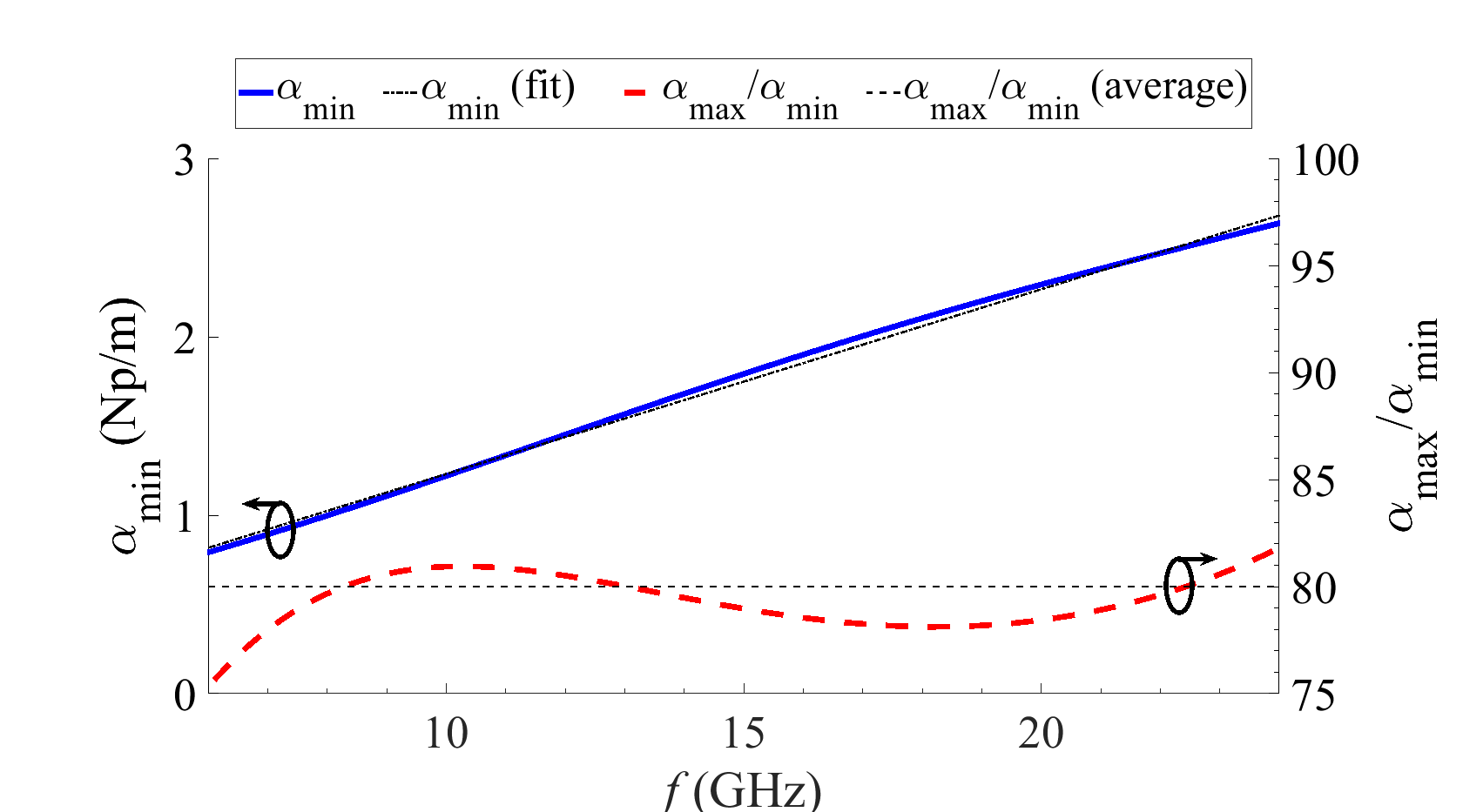}
\par\end{centering}
}\medskip{}
\subfloat[\label{fig:alpha_vs_tandelta}]{\noindent \begin{centering}
\includegraphics[width=0.9\columnwidth]{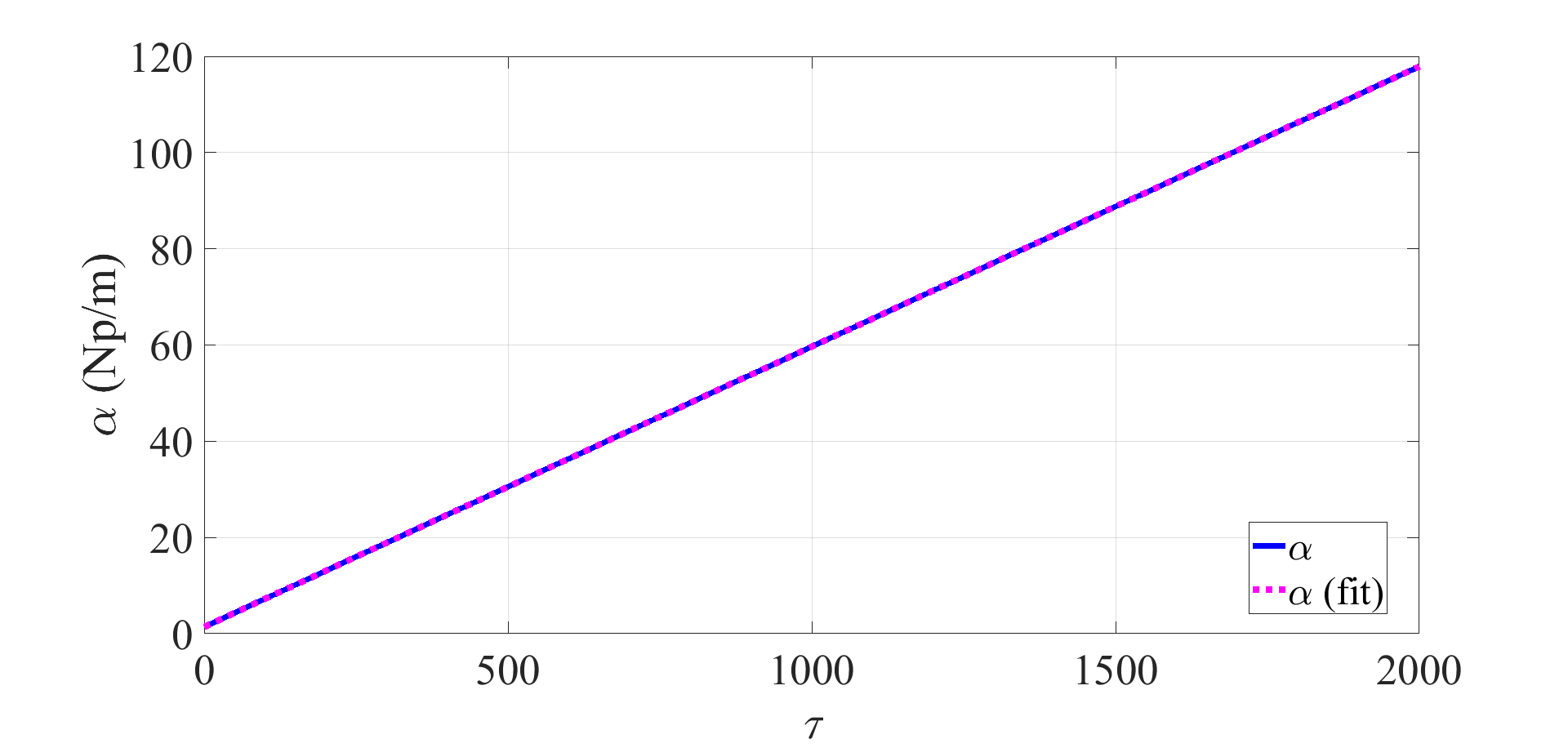}
\par\end{centering}
}\caption{(a) Attenuation coefficient, $\alpha_{\mathrm{min}}$, and ratio of
maximum attenuation coefficient to minimum attenuation coefficient
versus frequency for the helix SWS. The minimum attenuation coefficient
corresponds to the region of the SWS with dielectric rods without
loss coating, i.e., the attenuation at the input or output ports of
the TWT labeled in Fig. \ref{fig:unit_cell}, (b) Attenuation coefficient
versus dielectric loss tangent scaling factor at 12 GHz. Attenuation
is due to the lossy dielectric rods, tungsten helix, and copper outer
wall. The attenuation coefficient for the SWS is related to the frequency
by a first-order polynomial fitting (dotted black lines), as is the
relationship between the loss tangent scaling factor and attenuation
coefficient (magenta dashed line).}
\end{figure}

One can observe from Fig. \ref{fig:alpha_min_ratio} that the frequency-dependence
of $\alpha_{\mathrm{min}}$ can be fitted to a first-order polynomial
$\alpha_{\mathrm{min}}\approx0.1035f_{\mathrm{GHz}}+0.1961\ \mathrm{Np/m}$
and that the average value of $\alpha_{\mathrm{max}}/\alpha_{\mathrm{min}}$
is approximately 80 over the band of interest. Although the ratio
$\alpha_{\mathrm{max}}/\alpha_{\mathrm{min}}$ varies by as much as
$\pm6\%$ from the average value, this frequency-dependence does not
make a notable difference in the gain-vs-frequency plots. Using a
maximum attenuation coefficient of $\alpha_{\mathrm{max}}=84.8\alpha_{\mathrm{min}}$
or $\alpha_{\mathrm{max}}=75.2\alpha_{\mathrm{min}}$(i.e. $\pm6\%$
deviation from the average scaling factor of 80) did not alter the
shape of gain-vs-frequency profiles of our model or LATTE with respect
to PIC results, only their peak gain. Thus, for simplicity, we use
a constant scaling factor of $\alpha_{\mathrm{max}}=80\alpha_{\mathrm{min}}$in
our model. By scaling $\alpha_{\mathrm{min}}$ to obtain $\alpha_{\mathrm{max}}$,
we make the simplifying assumption that attenuation coefficients in
the low-loss and high-loss regions of the SWS have the same dependence
on frequency. For the frequency range of interest, this appears to
be a reasonable approximation. Additionally, we do not consider how
large losses affect the phase velocity and interaction impedance of
the cold structure in our above examples.

In Fig. \ref{fig:alpha_vs_tandelta} we show the attenuation coefficient
versus dielectric loss tangent, where the dielectric loss tangent
is related to the minimum loss tangent by the scaling factor, $\tau$,
as $\tan\delta=\tau\tan\delta_{\mathrm{min}}$. It is clear that the
attenuation coefficient is related to the loss tangent of our dielectric
support rods by a first order polynomial $\alpha\approx0.05825\tau+1.406\ \mathrm{Np/m}$.
In Fig. \ref{fig:alpha_vs_tandelta}, the vertical offset of the attenuation
coefficient when there is no dielectric loss tangent is due to metal
losses in the copper walls and tungsten helix. The first-order relationship
between the dielectric loss tangent and attenuation coefficient is
useful because the position-dependent attenuation profile will have
an identical shape to the position-dependent loss tangent profile
that we impose in full-wave PIC simulations.

\noindent 

\section{Sever Gap Transfer Matrix\label{sec:Sever-Gap-Tmatrix}}

\begin{figure}
\noindent \begin{centering}
\includegraphics[width=0.9\columnwidth]{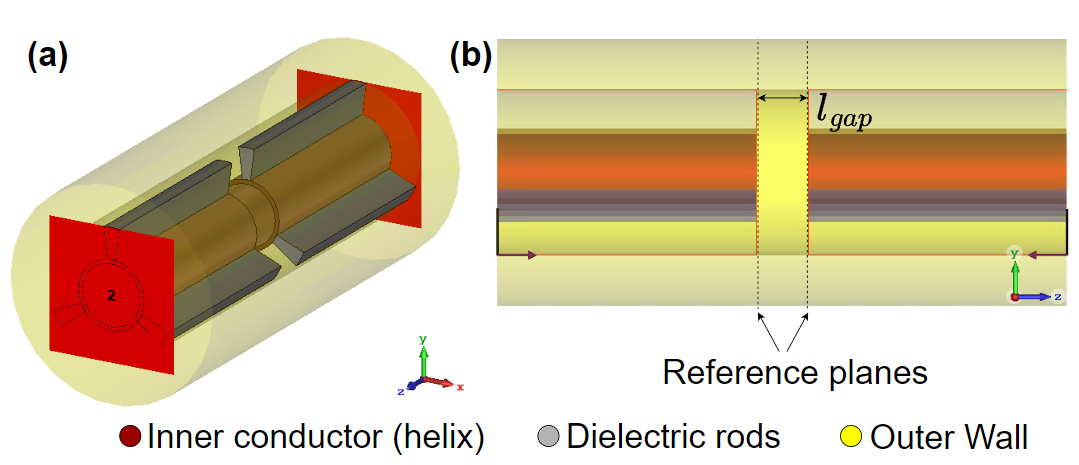}
\par\end{centering}
\caption{Model for determining equivalent capacitive pi network of a helix
sever gap. (a) Isometric view of CST model, and (b) side view of CST
model.\label{fig:sever_gap_model}}
\end{figure}

An imperfect sever, i.e. one that allows EM waves to weakly transmit
between stages separated by a gap, can be represented as a capacitive
pi network in the helix structure. This is similar to the classical
case of a discontinuity in the center conductor of a coaxial transmission
line \citep{whinnery1944coaxial,green1965numerical,dawirs1969equivalent},
except the center conductor is hollow and helical in shape for our
case. Due to this difference, it is necessary to simulate the sever
gap in a full-wave solver to approximate the gap capacitance. However,
it becomes necessary to simplify the geometry even further to excite
the proper modes and accurately compute the gap capacitance. In our
simplified full-wave model for the sever gap, we replace the center
helix with a hollow inner conductor of the same inner and outer radii,
as shown in Fig. \ref{fig:sever_gap_model}. This simplified model
allows TEM waves to be excited by both wave ports to the left and
right of the sever gap. The wave ports are both positioned 5 mm away
from the ends of the 1 mm sever gap, with reference planes on each
side of the inner conductor discontinuity. The disagreement between
our PIC results and our model for the two-stage TWT in Sec. \ref{sec:Results}
may partially be attributed to this simplified model for the sever
gap. A more complex model (with a tape helix instead of a hollow tube
for the center conductor) can be used in future works to accurately
model the sever gap, however it will be a significant challenge to
excite the proper helix modes and find the scattering parameters at
the reference planes of the gap.

Using full-wave simulations, we compute the scattering parameters
of the simplified two-port sever gap network and convert them to ABCD
parameters by using the helix characteristic impedance from Appendix
\ref{sec:helix_zc}. The equivalent frequency-dependent series and
shunt capacitances of the sever gap can then be calculated from the
forward ABCD parameters \citep[(Ch. 4)]{pozar2009microwave} as

\begin{equation}
\left\{ \begin{array}{l}
C_{1}=\frac{-1}{j\omega B}\\
C_{2}=AC_{1}-C_{1}
\end{array}\right.
\end{equation}

\begin{figure}
\noindent \begin{centering}
\includegraphics[width=0.9\columnwidth]{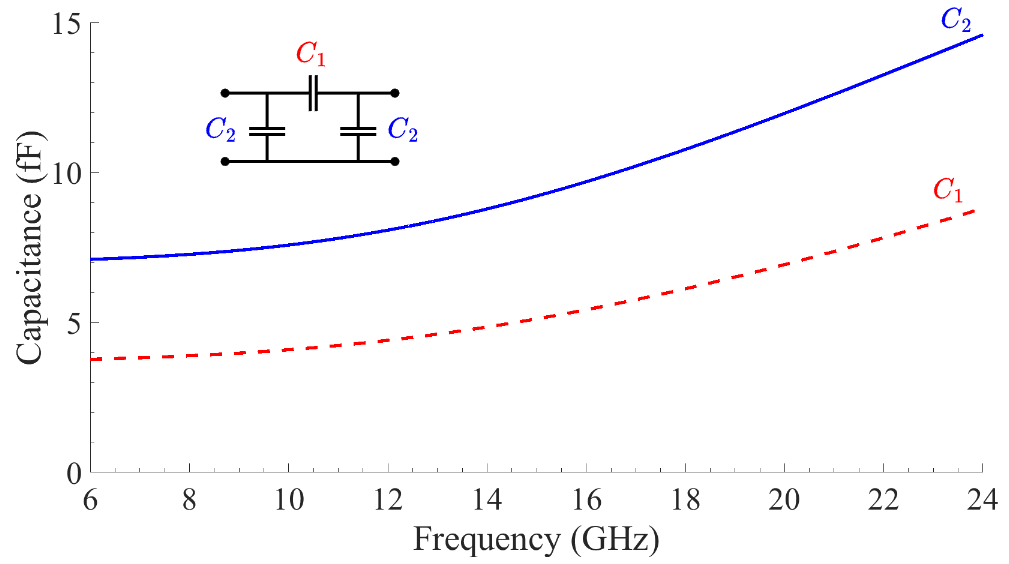}
\par\end{centering}
\caption{Equivalent two-port pi-network of helix sever gap and the corresponding
capacitance values versus frequency.\label{fig:Equivalent-C1C2}}
\end{figure}

Conversely, the two-port ABCD parameters of the capacitive pi network
shown in Fig. \ref{fig:Equivalent-C1C2} can be represented as

\begin{equation}
\left[\begin{array}{c}
V_{2}\\
I_{2}
\end{array}\right]=\underline{\boldsymbol{\mathrm{T}}}_{\mathrm{\mathrm{gap},c}}\left[\begin{array}{c}
V_{1}\\
I_{1}
\end{array}\right],
\end{equation}
where

\begin{equation}
\underline{\boldsymbol{\mathrm{T}}}_{\mathrm{\mathrm{gap},c}}=\left[\begin{array}{cc}
\frac{C_{1}+C_{2}}{C_{1}} & \frac{-1}{j\omega C_{1}}\\
\frac{-j\omega\left(2C_{1}C_{2}+C_{2}^{2}\right)}{C_{1}} & \frac{C_{1}+C_{2}}{C_{1}}
\end{array}\right].
\end{equation}

\section{Plasma Frequency Reduction Factor \label{sec:Plasma-Frequency-Reduction}}

For an electron beam moving at an average velocity $u_{0}$, these
plasma frequency oscillations result in propagating fast and slow
space charge waves (relative to the average electron velocity). The
plasma frequency of a linear electron beam of cross-sectional area,
$A$, is given as,

\begin{equation}
\omega_{\mathrm{p}}=\sqrt{-\frac{\rho_{0}\eta}{A\varepsilon_{0}}}=\sqrt{\frac{I_{0}u_{0}}{2V_{0}A\varepsilon_{0}}}.\label{eq:omega_p}
\end{equation}

\begin{figure}
\noindent \begin{centering}
\includegraphics[width=0.9\columnwidth]{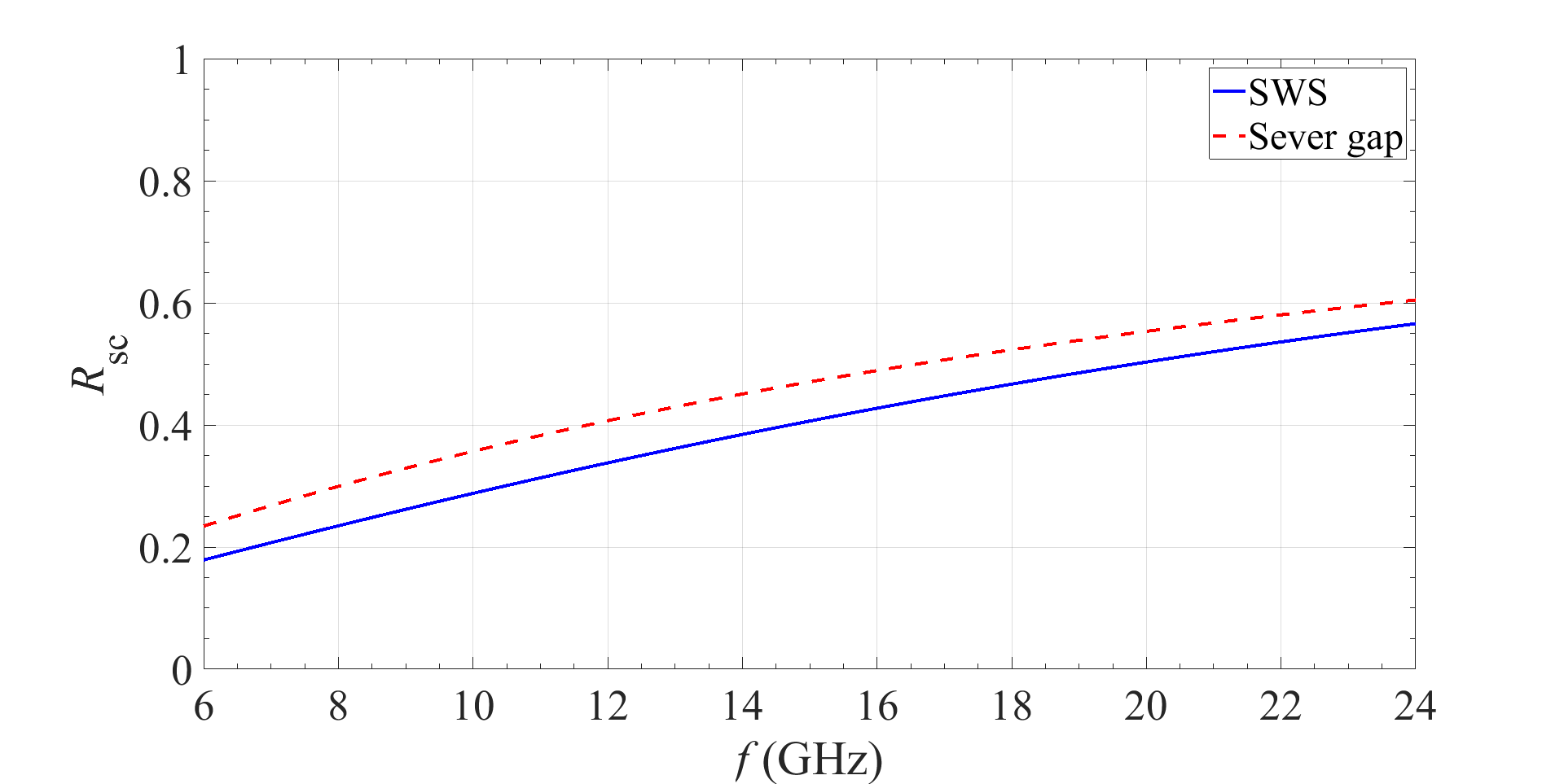}
\par\end{centering}
\caption{Plasma frequency reduction factor versus frequency in the SWS and
in the sever gap regions\label{fig:Rsc}}
\end{figure}

As explained in \citep{branch1955plasma,datta2009simple,ramo1939electronic},
the finite cross-section of the electron beam, along with surrounding
metallic walls will make the scalar electric potential of the electron
beam nonuniform over the beam cross-section. Because of this fact,
the plasma frequency of the beam will be reduced by a plasma frequency
reduction factor ($R_{\mathrm{sc}}$), $\omega_{\mathrm{q}}=R_{\mathrm{sc}}\omega_{\mathrm{p}}$.
The closed-form frequency-dependent value we use for $R_{\mathrm{sc}}$
is calculated from \citep{antonsen1998traveling,datta2009simple}
as,

\noindent 
\begin{equation}
R_{\mathrm{sc}}^{2}=1-2\mathrm{I_{1}}(\beta_{0}r_{\mathrm{b}})\left(\mathrm{K_{1}}(\beta_{0}r_{\mathrm{b}})+\frac{\mathrm{K_{0}}(\beta_{0}r_{\mathrm{h}})}{\mathrm{I_{0}}(\beta_{0}r_{\mathrm{h}})}\mathrm{I_{1}}(\beta_{0}r_{\mathrm{b}})\right),\label{eq:Rsc}
\end{equation}

\noindent where, we assume the beam has a cylindrical cross-section
with radius $r_{\mathrm{b}}$ and the helix is approximated as a metallic
cylinder with inner radius $r_{\mathrm{h}}$. Furthermore, $\mathrm{I}$
and $\mathrm{K}$ are modified Bessel functions of the first and second
kind, respectively. A formulation for $R_{\mathrm{sc}}$for the case
of an electron beam confined by a helical slow-wave structure, which
is approximated using the sheath helix model, has also been provided
in \citep{antonsen1998traveling}, though it is not as trivial to
use. In the sever gap region of the multi-stage TWT, we calculate
$R_{\mathrm{sc}}$ from Eqn. (\ref{eq:Rsc}) like above, but with
the wall radius $r_{\mathrm{w}}$ substituted in place of $r_{\mathrm{h}}$.
The reduction factor in both the SWS and the sever gap region is plotted
versus frequency in Fig. \ref{fig:Rsc}.

\section{Hot Dispersion Relation from System Matrix in a Uniform TWT\label{sec:Dispersion-Relation-from}}

In the supplementary material of \citep{rouhi2021exceptional}, it
was shown that the modal solutions for a hot TWT in the form $\mathbf{\Psi}(z)=\mathbf{\Psi}e^{-jkz}$
can be found from the dispersion equation $\mathrm{Det}(\mathrm{\underline{\mathbf{M}}-\mathit{k}\underline{\mathbf{I}}})=0$
that matches the fourth-order dispersion relation from Pierce theory
\citep[Ch. 2]{pierce1950traveling1}\citep[Ch. 8]{tsimring2006electron},
where $\underline{\mathbf{I}}$ is a $4\times4$ identity matrix and
$k$ is the hot complex propagation constant of the TWT system. Like
in \citep{rouhi2021exceptional}, the dispersion relation for a lossy
TWT is calculated from the determinant of the system matrix in Eqn.
(\ref{eq:system_matrix}) to demonstrate that it is the same as the
one found in conventional Pierce theory \citep[Ch. 2]{pierce1950traveling1}
\citep[Ch. 8]{tsimring2006electron}.

The fourth-order dispersion equation for the lossy TWT is found from
the determinantal equation above as

\begin{equation}
\left(k^{2}-k_{\mathrm{c}}^{2}\right)\left(\left(k-\beta_{0}\right)^{2}-\zeta_{\mathrm{sc}}g\right)=-a^{2}gZ_{\mathrm{c}}k_{\mathrm{c}}k^{2}\label{eq:dispersion_relation}
\end{equation}

\noindent where $k_{\mathrm{c}}$ has complex value and $Z_{\mathrm{c}}$is
assumed to be real under the low-loss approximation discussed in Sec.
\ref{sec:Transmission_line_model}. The Pierce gain parameter is defined
as $C^{3}=a^{2}Z_{\mathrm{c}}I_{0}/(4V_{0})$ (where the term $a^{2}Z_{\mathrm{c}}$
has been substituted for $Z_{\mathrm{P}}$) and the plasma propagation
constant is defined as $\beta_{q}=\omega_{\mathrm{q}}/u_{0}$. Substituting
the Pierce gain parameter and plasma propagation constant into (\ref{eq:dispersion_relation}),
we obtain the conventional Pierce dispersion relation

\begin{equation}
\left(k^{2}-k_{\mathrm{c}}^{2}\right)\left(\left(k-\beta_{0}\right)^{2}-\beta_{\mathrm{q}}^{2}\right)=-2C^{3}k_{\mathrm{c}}\beta_{0}k^{2}.\label{eq:dispersion_relation_pierce}
\end{equation}

If there are no losses (i.e. $k_{\mathrm{c}}=\beta_{\mathrm{c}}$),
the dispersion relation in Eqns. (\ref{eq:dispersion_relation}) and
(\ref{eq:dispersion_relation_pierce}) match those found in the supplementary
material of \citep{rouhi2021exceptional}. Furthermore, if one assumes
that $k\approx k_{\mathrm{c}}$ like was done by Pierce, then $k^{2}-k_{\mathrm{c}}^{2}\approx2k_{\mathrm{c}}(k-k_{\mathrm{c}})$
and the fourth-order modal dispersion relation in Eqn. (\ref{eq:dispersion_relation_pierce})
is reduced to a third-order dispersion relation which neglects the
backward-propagating wave supported by the TWT and only considers
three forward-propagating waves \citep[Ch. 2]{pierce1950traveling1}\citep[Ch. 8]{tsimring2006electron}

\begin{equation}
(k-k_{\mathrm{c}})\left(\left(k-\beta_{0}\right)^{2}-\beta_{\mathrm{q}}^{2}\right)=-C^{3}k_{\mathrm{c}}^{2}\beta_{0}.\label{eq:dispersion_relation_pierce_3wave}
\end{equation}

This modal dispersion relation in Eqn. (\ref{eq:dispersion_relation_pierce_3wave})
above is often used in conventional Pierce theory. A comparison between
our generalized Pierce model (with considers all four waves in the
TWT) and the conventional Pierce model (which considers only three
waves) is shown for a homogeneous TWT example in Appendix \ref{sec:Homogeneous-TWT-Example}.

\section{Spatially Homogeneous TWT Example\label{sec:Homogeneous-TWT-Example}}

To further validate our model, we provide a comparison between our
model and the conventional three-wave model provided by Pierce \citep[Ch. 9]{pierce1950traveling1,pierce1951waves,pierce1947theoryTWT}\citep[Ch. 8]{tsimring2006electron},
in terms of (i) the gain vs frequency, and (ii) RF power vs longitudinal
distance, for the restrictive case of a simple, spatially homogeneous,
dispersive, and lossy, single-stage TWT. The conventional Pierce model
considers only three propagating hot waves in the TWT system and it
neglects the backward wave supported by the TWT, whereas our model
considers all four hot waves. The power gain of the conventional Pierce
model is calculated as $G_{\mathrm{dB}}(z)=20\mathrm{log}_{10}\left|E(z)/E_{0}\right|$,
where the position-dependent longitudinal electric field along the
TWT from \citep[Eqn. 3.8]{pierce1951waves} is $E(z)=\frac{1}{3}E_{0}e^{-j\beta_{0}z}(e^{\delta_{1}z}+e^{\delta_{2}z}+e^{\delta_{3}z})$,
$E_{0}$ is the longitudinal electric field at the start of the TWT,
and the incremental propagation constant for each wave (subscripts
1, 2, 3) is $\delta=-j(k-\beta_{0})/(\beta_{0}C)$, which is determined
from the three e wavenumber solutions $k=k_{1},k_{2},k_{3}$, respectively,
that are calculated from the third-order dispersion relation shown
in Eqn. (\ref{eq:dispersion_relation_pierce_3wave}). With conventional
Pierce theory, the gain of a finite-length TWT of length $L$ is simply
$G_{\mathrm{dB}}(L)$.

In this example, the beam parameters, TWT length $L$, SWS dimensions,
and input RF power are the same as described for the single-stage
TWT example that was shown in Sec. \ref{sec:Results}, with the difference
that in this appendix only the intrinsic attenuation coefficient (i.e.
the one associated to $\alpha_{\mathrm{min}}$) is used over the length
of the TWT, which is frequency dependent but not position dependent,
in both the conventional three-wave Pierce theory and our model (referred
to as the generalized Pierce model in this paper). Furthermore, in
both models, we consider the same dispersive parameters described
in this work (such as wavenumber, attenuation coefficient, reduced
plasma frequency, interaction impedance, characteristic impedance,
and coupling coefficient $a$).

\begin{figure}
\noindent \begin{centering}
\includegraphics[width=0.9\columnwidth]{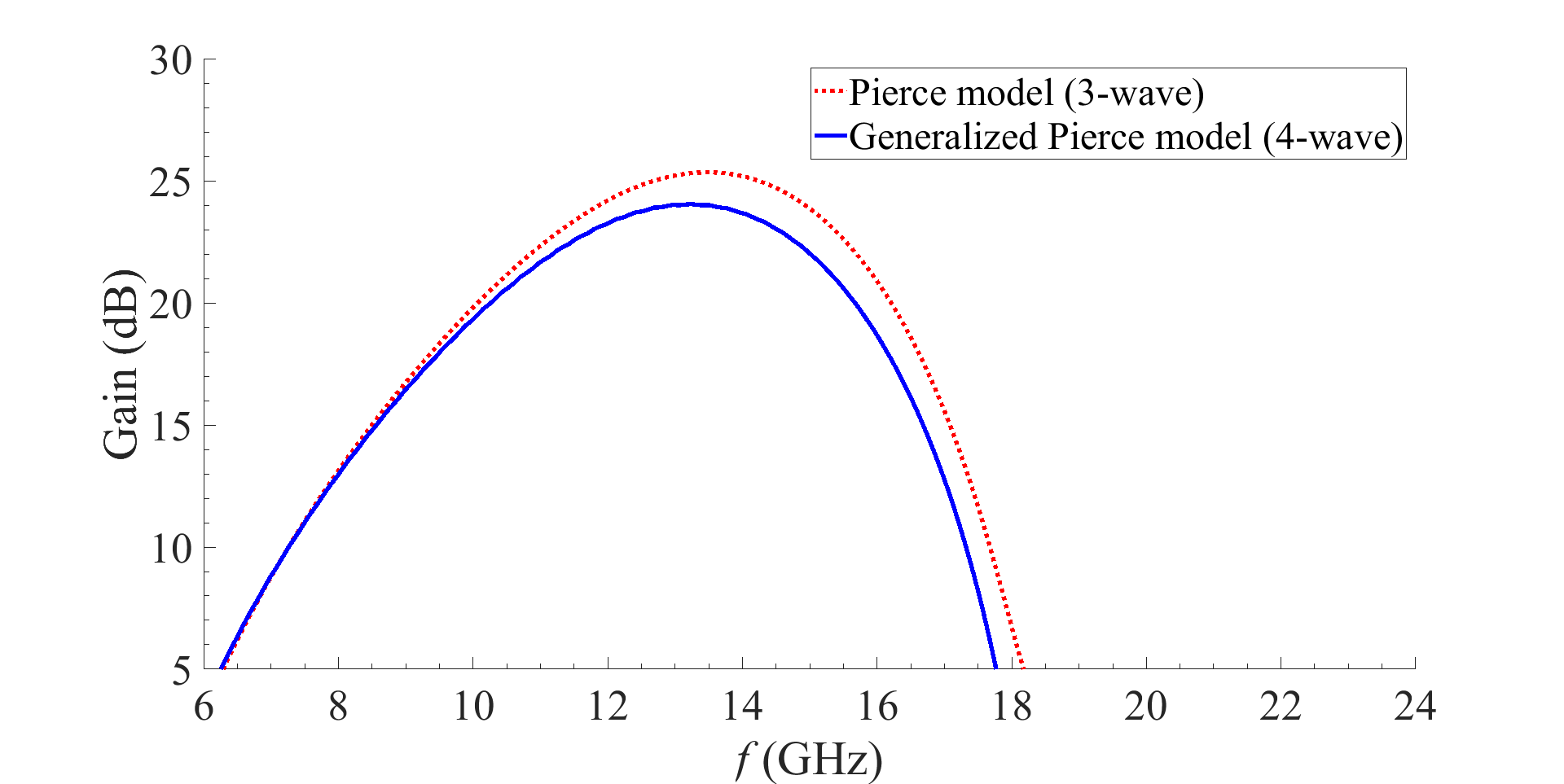}
\par\end{centering}
\caption{Gain vs frequency for a spatially homogeneous single-stage TWT using
the generalized Pierce model discussed in this paper (solid blue),
compared to that from Pierce's conventional three-wave model (dotted
red). \label{fig:homogeneous-gain-vs-frequency}}
\end{figure}

We note that, unlike our generalized Pierce model, the conventional
three-wave Pierce model that we use for comparison does not include
a coupling parameter and instead assumes the characteristic impedance
of the SWS to be equal to the interaction impedance $Z_{\mathrm{P}}$
(hence it assumes $a=1$), and both ends of the TWT are terminated
with $Z_{\mathrm{P}}$. In our model, the single-stage TWT is terminated
on both ends with its cold characteristic impedance $Z_{\mathrm{c}}$
and its gain vs frequency is plotted in Fig. \ref{fig:homogeneous-gain-vs-frequency}.
Since there is no significant loss patterning in the center of the
single-stage TWT as there was in the example in Sec. \ref{sec:Results},
the peak gain is approximately 16 dB higher than the one found in
Fig. \ref{fig:single_stage_gain_vs_freq}. Furthermore, the peak gain
calculated from the conventional Pierce model is approximately 1.5
dB higher than the gain obtained from our generalized Pierce model
that considers all four waves.

\begin{figure}
\noindent \begin{centering}
\includegraphics[width=0.9\columnwidth]{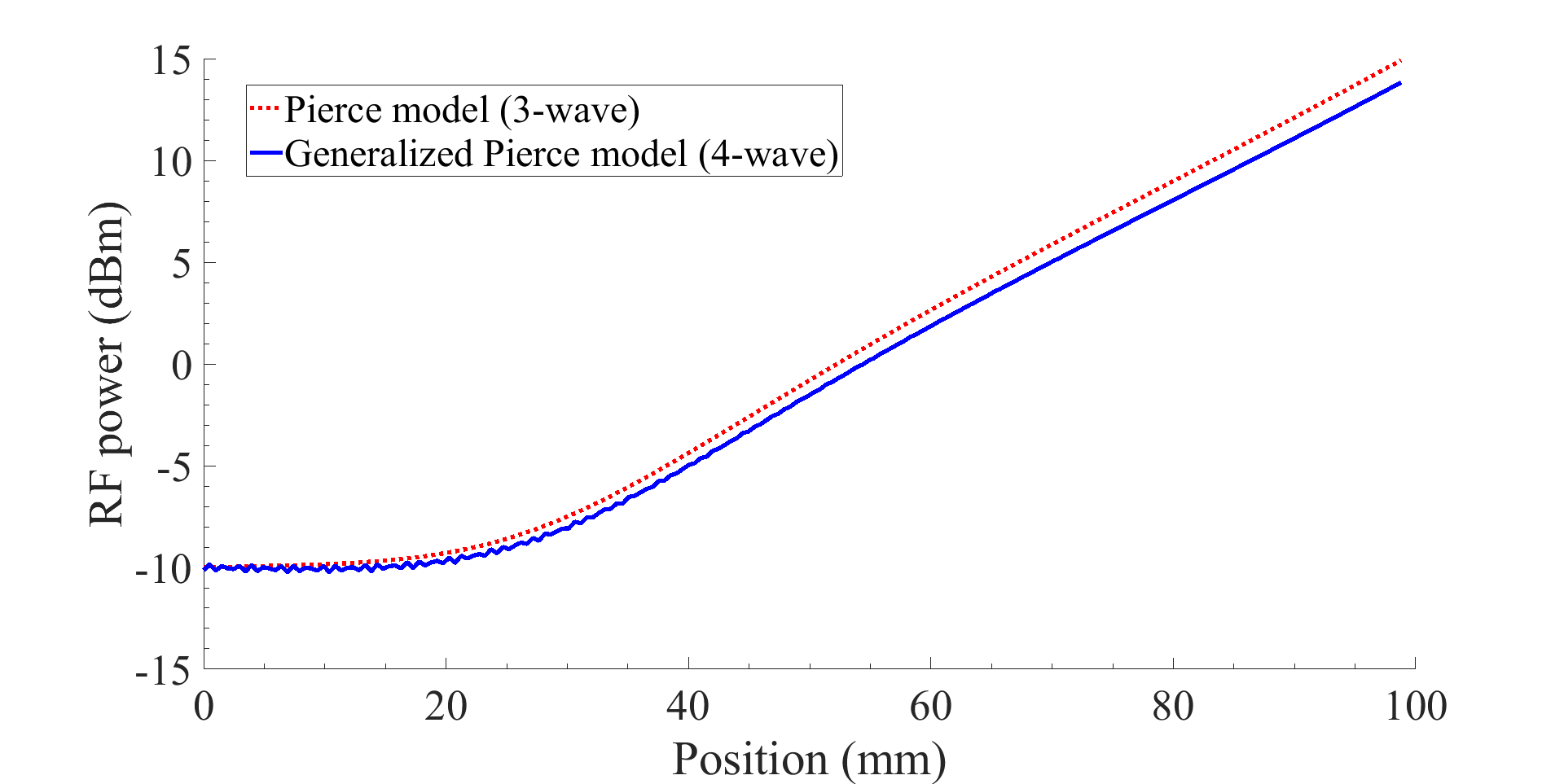}
\par\end{centering}
\caption{RF power vs position at 12 GHz for a spatially homogeneous single-stage
TWT, calculated using our generalized Pierce model (solid blue), and
Pierce's conventional three-wave model (dotted red). \label{fig:homogeneous-pwr-vs-z}}
\end{figure}

The RF power from our generalized Pierce model, $P(z)=\frac{1}{2}\mathrm{Re}\left[V(z)I(z)^{*}\right]$
obtained by using $V(z)$ and $I(z)$ of the state vector, is plotted
against position for this spatially homogeneous example in Fig. \ref{fig:homogeneous-pwr-vs-z}.
Whereas, for the conventional three-wave Pierce model, the position-dependent
power is calculated from the position-dependent power gain as $P(z)=P_{\mathrm{in,dBm}}+G_{\mathrm{dB}}(z)$
and is also plotted in Fig. \ref{fig:homogeneous-pwr-vs-z}, where
the input power for both cases is $P_{\mathrm{in,dBm}}=-10~\mathrm{dBm}$.
The RF power at the collector-end of the TWT calculated from conventional
Pierce theory is approximately 1 dBm higher than the power calculated
by our model at 12 GHz.

\section{beam-wave power conservation\label{sec:energy-conservation-proof}}

Referring to the homogeneous small-signal TL model shown in Fig. \ref{fig:TL_model}(b),
using simple TL theory or Lagrangian theory, it is easy to show that
that the time-average power flowing in the TL is $P_{\mathrm{TL}}=\frac{1}{2}\mathrm{Re}\left(VI^{*}\right).$
In a lossless TL, the power variation $dP_{\mathrm{TL}}/dz$ is equal
to the power transfer per-unit-length from the dependent current generator
$i_{\mathrm{s}}=-a\left(dI_{\mathrm{b}}/dz\right)$ (representing
the effect of electron beam to the guided RF wave) \citep[p. 309]{hutter1960beam}\citep[p. 286]{figotin2021analytic}\citep{tamma2014extension},
i.e.,

\begin{equation}
\frac{dP_{\mathrm{TL}}}{dz}=\frac{1}{2}\mathrm{Re}\left(Vi_{\mathrm{s}}^{*}\right)=-\frac{1}{2}|V|^{2}\mathrm{Re}\left(Y_{\mathrm{b}}^{*}\right),\label{eq:dP1-2}
\end{equation}

\noindent where $*$ denotes complex conjugation, and we have used
the concept of electronic beam admittance per unit length $Y_{\mathrm{b}}=-i_{\mathrm{s}}/V$
given in \citep{rouhi2023parametric,tamma2014extension}. In lossy
TLs, there is an additional term that represents the attuenaution
along the TL. The time-average power transfer per-unit-length of TWT
from the electron beam (considered as an impressed ac current source
$I_{\mathrm{b}}$) to the guided RF wave, from \citep[p. 359]{gewartowski1965principles}\citep[p. 324]{tsimring2006electron}\citep[p. 168]{figotin2021analytic},
is

\begin{equation}
\frac{dP_{\mathrm{b}}}{dz}=\frac{1}{2}\mathrm{Re}\left(E_{z}I_{\mathrm{b}}^{*}\right),\label{eq:dP2}
\end{equation}

\noindent where the axial electric field of the guided RF wave that
couples to the electron beam is $E_{z}=-a(dV/dz).$ Considering the
propagation constant $k$ of the hot amplifying mode, we obtain the
expressions $E_{z}=jakV$ and $I_{\mathrm{b}}=ji_{\mathrm{s}}/(ka)$.
Substituting these expressions for $E_{z}$ and $I_{\mathrm{b}}$
into Eqn. (\ref{eq:dP2}), we obtain

\begin{equation}
\frac{dP_{\mathrm{b}}}{dz}=\frac{1}{2}\mathrm{Re}\left(Vi_{\mathrm{s}}^{*}\frac{ka}{(ka)^{*}}\right)=-\frac{1}{2}|V|^{2}\mathrm{Re}\left(Y_{\mathrm{b}}^{*}\frac{ka}{(ka)^{*}}\right).\label{eq:dP2-1-1}
\end{equation}

When the beam-wave coupling coefficient $a$ is real (as considered
in this paper) and $|\mathrm{Re}(k)|\gg|\mathrm{Im}(k)|$, the time-average
powers per unit length are nearly equal, i.e., $dP_{\mathrm{TL}}/dz\approx dP_{\mathrm{b}}/dz$.
There is a slight inconsistency in the power of the guided RF wave
calculated using Eqn. (\ref{eq:dP2-1-1}), compared to Eqn. (\ref{eq:dP1-2}).
However, the inconsistency between powers is negligible in the small-signal
regime, since $\mathrm{Re}(k)$ is typically one to two orders of
magnitude larger than $\mathrm{Im}(k)$, as demonstrated in \citep{rouhi2021exceptional,rouhi2023parametric,wong2018modification}.
This inconsistency in calculated powers is intrinsic to the Pierce
model that we base our model upon \citep{pierce1947theoryTWT,pierce1950traveling1,pierce1951waves}
(the Pierce model assumes that $a=1$). However, it has been demonstrated
in \citep[Ch. 22]{figotin2021analytic}\citep{figotin2021generalized}
that the Pierce model remains energetically self-consistent from the
Lagrangian point of view.

\bibliographystyle{aipnum4-1}

\end{document}